\documentclass[12pt]{article}
\usepackage{tgpagella}
\usepackage[utf8]{inputenc}
\usepackage{color}
\usepackage{float}
\usepackage{amsmath}
\usepackage{amssymb}
\usepackage{graphicx}
\usepackage{geometry}
\usepackage{setspace}
\usepackage[authoryear]{natbib}
\usepackage[bookmarks=false,
 breaklinks=false,pdfborder={0 0 1},backref=section,colorlinks=true]
 {hyperref}
\hypersetup{
 pagebackref=true,linkcolor=blue,filecolor=magenta,urlcolor=blue,citecolor=blue}

\makeatletter

\DeclareTextSymbolDefault{\textquotedbl}{T1}

\usepackage{amsfonts}\usepackage{eurosym}\usepackage{ulem}\usepackage{caption}\usepackage{color}\usepackage{sectsty}\usepackage{comment}\usepackage{footmisc}\usepackage{pdflscape}\usepackage{array}\usepackage[dvipsnames]{xcolor}
\usepackage[toc,page]{appendix}
\usepackage{adjustbox}
\usepackage{xcolor}
\usepackage{soul}

\newcolumntype{L}[1]{>{\raggedright\let\newline\\arraybackslash\hspace{0pt}}m{#1}}
\newcolumntype{C}[1]{>{\centering\let\newline\\arraybackslash\hspace{0pt}}m{#1}}
\newcolumntype{R}[1]{>{\raggedleft\let\newline\\arraybackslash\hspace{0pt}}m{#1}}

\definecolor{tt}{HTML}{b52a1d}

\ifdefined\showcaptionsetup
\fi
\emergencystretch=\maxdimen
\ifdefined\showcaptionsetup
 \PassOptionsToPackage{caption=false}{subfig}
\fi
\usepackage{subfig}
\makeatother

\begin{document}
\begin{titlepage} 
\title{A Millennium of UK Business Cycles: Insights from Structural VAR Analysis\thanks{We thank participants of the Workshop in Empirical and Theoretical
Economics at King's College London for their useful comments. The
views expressed in this paper are those of the authors and do not
represent those of the Bank for International Settlements, the Central
Bank of Brazil, or any of their Committees.}}
\author{Leonardo N. Ferreira \thanks{Central Bank of Brazil. Email: leonardo.ferreira@bcb.gov.br}
\and Haroon Mumtaz \thanks{Queen Mary University of London. Email: h.mumtaz@qmul.ac.uk }
\and Gabor Pinter \thanks{Bank for International Settlements. Email: gabor.pinter@bis.org }}
\date{\today}

\maketitle
\singlespacing \singlespacing %\vspace{-1.9cm}
\begin{abstract}
\noindent We study macroeconomic fluctuations in the United Kingdom
over seven centuries (1271--2022) using a time-varying VAR with stochastic
volatility. We identify business cycle shocks as innovations explaining
the largest share of future output variance. Before 1900, these shocks
display a stagflationary, supply-driven pattern, while post-1900 shocks
become demand-driven, raising both output and inflation. Output volatility
declines over time, peaking in the seventeenth century. Monetisation
had large real effects in the sixteenth and seventeenth centuries,
shifting to more inflationary impacts thereafter. Our results highlight
how business cycle dynamics evolve with institutional, monetary, and
structural transformations. 
\end{abstract}
\vspace{0.5cm}

 \vspace{0in}

\noindent %\begin{center}
%\Large	
%{Preliminary draft. Please do not quote.}
%\end{center}
\setcounter{page}{0} \thispagestyle{empty} \end{titlepage} \pagebreak\newpage{}

\doublespacing

\newpage{}

\section{Introduction}

Macroeconomic fluctuations have been a persistent feature of human
economic history. Yet, modern empirical business cycle research typically
restricts attention to the post-World War II era, often due to data
availability and methodological convenience. As a result, our understanding
of macroeconomic dynamics is largely derived from a relatively short
and historically exceptional period.

In this paper, we take a long-run perspective and study the evolution
of macroeconomic fluctuations in the United Kingdom over more than
seven centuries, from 1271 to 2022. By combining historical macroeconomic
data on output, inflation, and money with a structural time-varying
vector autoregression (TVP-VAR) framework that accounts
for measurement error in long-run macroeconomic series, we provide
novel evidence on the changing nature of business cycles.

The long time span allows us to investigate a fundamental question:
have the sources and propagation mechanisms of business cycles changed
over time? Drawing inspiration from recent work by \citet{angeletos2020},
we identify the main business cycle shock in each year as the innovation
that explains the largest share of short-term GDP growth variance.
This agnostic approach avoids imposing modern structural interpretations
on earlier eras. We estimate a time-varying VAR with stochastic volatility
and fat-tailed shocks, enabling us to capture drifting macroeconomic
regimes, levels of uncertainty, and non-Gaussian features that characterize
pre-modern economies. Our analysis yields five main results.

First, our findings reveal a striking historical transformation in
the nature of business cycle shocks. Prior to 1900, macroeconomic
fluctuations were dominated by supply-side disturbances, with negative
shocks reducing output per capita and raising prices -- a stagflationary
pattern consistent with historical famines, wars, and climate events.
In contrast, from the early twentieth century onward, business cycle
shocks take on a demand-driven character, producing positive co-movement
between output and inflation. This regime shift coincides with the
rise of fiat money, modern central banking, and active fiscal-monetary
stabilization policies (\citealp{rogoffcarmern2008,rogoff2013}).

Second, our results reveal a long-run decline in output volatility,
with the 17th century marking the peak of economic instability over
the past millennium due to wars, political upheaval, and climatic
shocks (\citealp{parker2013}). Volatility was lower both before and
after this period. Even the turbulent business cycles of the twentieth
century show smaller output fluctuations than the 17th century, except
for the temporary spike during COVID-19. Overall, this pattern suggests
a centuries-long Great ``Moderation'' (\citealp{Kim1999,McConnell2000,prim2005}),
as economic development and stabilising institutions gradually reduced
the amplitude of business cycle fluctuations.

Third, during the sixteenth and seventeenth centuries, money supply
variation had substantial real effects, consistent with historical
accounts that monetisation facilitated market transactions and economic
expansion (\citealp{palma2018modernisation,Palma2018supply,palma2022}).
However, while money variation explained a significant share of business
cycle fluctuations in earlier centuries, its influence became increasingly
inflationary from the eighteenth century onward, consistent with a
transition towards demand-dominated cycles.

Fourth, our sign-restricted VAR results for the post-1700 period show
that aggregate demand disturbances capture the effects of both pure
demand shocks and money supply shocks. While pure demand shocks account
for a slightly larger share of output and price variation, money supply
shocks also play a significant role, particularly in explaining fluctuations
in money itself.

Fifth, we find that the shift toward inflationary business cycle shocks
is more closely linked to the expansion of base money than to financial
deepening per se. While the M4/M0 ratio -- our proxy for financial
development -- rises sharply after 1970, the demand-type characteristics
of business cycle shocks had already emerged during the late nineteenth
century.

By tracing the evolution of British business cycles from the medieval
period to the present, our paper highlights the importance of long-run
historical data for macroeconomic analysis. We demonstrate that business
cycle dynamics are not stable or universal, but instead reflect the
institutional, monetary, and structural environment in which they
unfold. These insights are particularly relevant for interpreting
recent macroeconomic shocks -- such as COVID-19 and inflation surges
-- within a broader historical context.

\paragraph{Related Literature}

This paper primarily contributes to the literature that seeks the
drivers of business cycles. Empirical studies of modern business cycles
argue that business cycles are primarily driven by demand shocks,
thereby challenging the more traditional view from Real Business Cycle
(RBC) models (\citealt{Kydland1982}). The finding emerges both from
VAR-type models (\citealt{shapiro1988,Blanchard1989,gali1999aer})
as well as from more structural models (\citealt{Smets2007,Christiano2014}).
More recently, the findings of \citet{angeletos2020} support the
existence of a main business-cycle driver which fits the notion of
an aggregate demand shock. Our work extends this analysis to multiple
centuries, showing how the main business-cycle shock has changed from
supply-like to demand-like around 1850-1900.

Our analysis of a nearly millennium long dataset for the UK is inspired
by several recent papers (\citealp{hansen2020mill,Schmelzing2020boe,nason2023,rogoff2024AER,nakamura2025qje}).
What distinguishes our work is that we model the joint behaviour of
price and quantity variables, allowing us to distinguish between supply-
and demand-type shocks in the spirit of modern structural macroeconomic
models.\footnote{For example, \citet{nason2023} analysis inflation dynamics in a univariate
framework, \citet{rogoff2024AER} focuses on the long-run dynamics
of interest rates, and \citet{nakamura2025qje} focuses on real variables
and does not study inflation dynamics.}

\section{Data}

Our dataset is annual, spanning the period from 1271 to 2022.\footnote{As described in the appendix, we use the first 50 observations as
training sample.} Our source is "A millennium of macroeconomic data", compiled by
\citet{Ryland2018}. This database contains a broad set of macroeconomic
and financial data for the UK, in some cases, from 1086 to 2016. We
select the three variables that best characterise the macroeconomic
dynamics throughout the centuries and update them up to 2022. These
are GDP per capita, CPI, and money (M0) per capita.

\section{Empirical Model}

We use a time-varying VAR with stochastic volatility and fat tails
to model the dynamics of these variables: 
\begin{equation}
Y_{t}=c_{t}+\sum_{j=1}^{P}b_{j,t}Y_{t-j}+\Sigma_{t}^{1/2}e_{t},\ \ \ e\sim N(0,1),
\end{equation}
where $Y_{t}=\{\Delta y_{t},\Delta p_{t},\Delta m_{t}\}$ and $L=2$
represents the lag length. Following \citet{Cogley_Sargent_2005}
and \citet{prim2005}, the coefficients are time-varying with $\Phi_{t}=\Phi_{t-1}+Q^{1/2}\eta_{t}$,
where $\Phi_{t}=vec\left(\left[c_{t},b_{j,t}\right]\right)$ represents
the time-varying coefficients stacked in one vector and $\eta_{t}$
is a conformable vector of innovations. As in \citet{prim2005}, the
covariance matrix of the innovations $v_{t}$ is factored, $\Sigma_{t}=A_{t}^{-1}H_{t}A_{t}^{-1^{\prime}}$,
where the non-zero and non-one elements of $A_{t}$ follow a random
walk, $\alpha_{ij,t}=\alpha_{ij,t-1}+Q_{\alpha}^{1/2}v_{t}$. As we
also model fat tails, $H_{t}$ is a diagonal matrix with elements
$\sigma_{i,t}^{2}\frac{1}{\lambda_{i,t}}$ for $i=1,2,3$. Here, $\ln\sigma_{i,t}^{2}=\ln\sigma_{i,t-1}^{2}+g_{i}^{1/2}z_{t}$
and $\lambda_{i,t}$ are weights that lead to a scale mixture of normals
such that the orthogonalised residuals $A_{t}u_{t}\sim T\left(0,\sigma_{i,t},v_{\lambda,i}\right)$.
As shown by \citet{geweke1993} this scale mixture can be obtained
if one assumes that $p\left(\lambda_{k}\right)=\prod\limits_{i=1}^{T}\Gamma\left(1,v_{\lambda,k}\right)$.

Modelling fat tails is particularly important in our context, as it
provides robustness against extreme observations that may arise from
measurement error or archival uncertainty in pre-modern macroeconomic
data (\citealp{muller2025}). Rather than treating these outliers
as structural shifts, the t-distribution allows the model to downweight
their influence without distorting the underlying signal.\footnote{This feature is also valuable for handling modern data outliers, such
as those arising during the Covid-19 pandemic.}

\subsection{Identification}

Building on \citet{uhlig2004moves,Barsky2011,otrok2013,kurmann2021s,Chahrour2023},
we identify the business cycle shock as the VAR innovation that makes
the largest contribution to the forecast error variance (FEV) of $\Delta y_{t}$
at horizon 1. \citet{angeletos2020} show this is equivalent to identifying,
in the frequency domain, the shock that dominates the business-cycle
frequencies (1.5-8 years).

We decompose $\Sigma$ such that $\Omega\Omega'=\Sigma$ where $\Omega$
represents a contemporaneous impact matrix. $\Omega$ can be written
as $\Omega=\tilde{\Omega}Q$ where $Q$ is an orthonormal matrix that
rotates $\tilde{\Omega}$, the Cholesky decomposition of $\Sigma$.
The structural moving average representation of the VAR model is then
given by: 
\begin{equation}
Y_{t}=B(L)\Omega\varepsilon_{t},
\end{equation}
where $\varepsilon_{t}$ denotes the orthogonal shocks.

The k-period-ahead forecast error of the $i$th variable is defined
as 
\begin{equation}
Y_{it+k}-\hat{Y}_{it+k}=e_{1}\left[\sum_{j=0}^{k-1}B_{j}\tilde{\Omega}Q\varepsilon_{t+k-j}\right],
\end{equation}
where $e_{1}$ is a selection vector that picks out in the set of
variables. The proposed identification scheme thus amounts to finding
the column of Q that solves the following maximization problem: 
\begin{equation}
\arg\max_{Q_{1}}e_{1}'\left[\sum_{k=0}^{K}\sum_{j=0}^{k-1}B_{j}\tilde{\Omega}Q_{1}Q_{1}'\tilde{\Omega}'B_{j}'\right]e_{1},
\end{equation}
such that $Q_{1}'Q_{1}=1$, where $Q_{1}$ is the column of $Q$ that
corresponds to the shock that explains the largest proportion of the
FEV of the first variable in the VAR: $\Delta y_{t}$.

\section{Results}

\begin{figure}[bp!]
\caption{The effects of business cycle shocks over centuries}
\label{fig:IRFs-3} 
\begin{centering}
\subfloat[Impulse response functions\label{fig:IRFs}]{\begin{centering}
\includegraphics[width=1\textwidth]{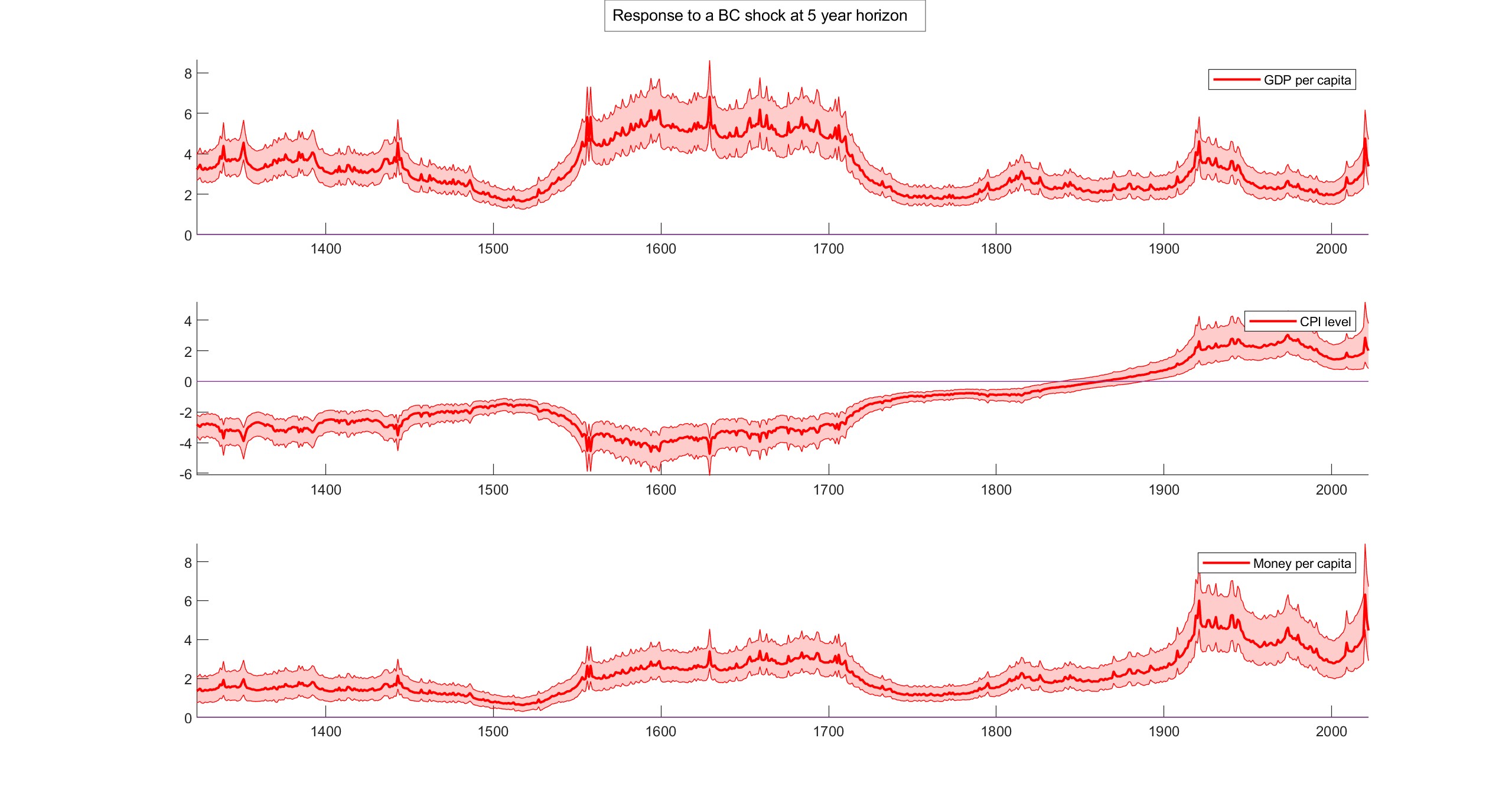} 
\par\end{centering}
}
\par\end{centering}
\begin{centering}
\subfloat[Forecast error variance decomposition\label{fig:IRFs-2}]{\begin{centering}
\includegraphics[width=1\textwidth]{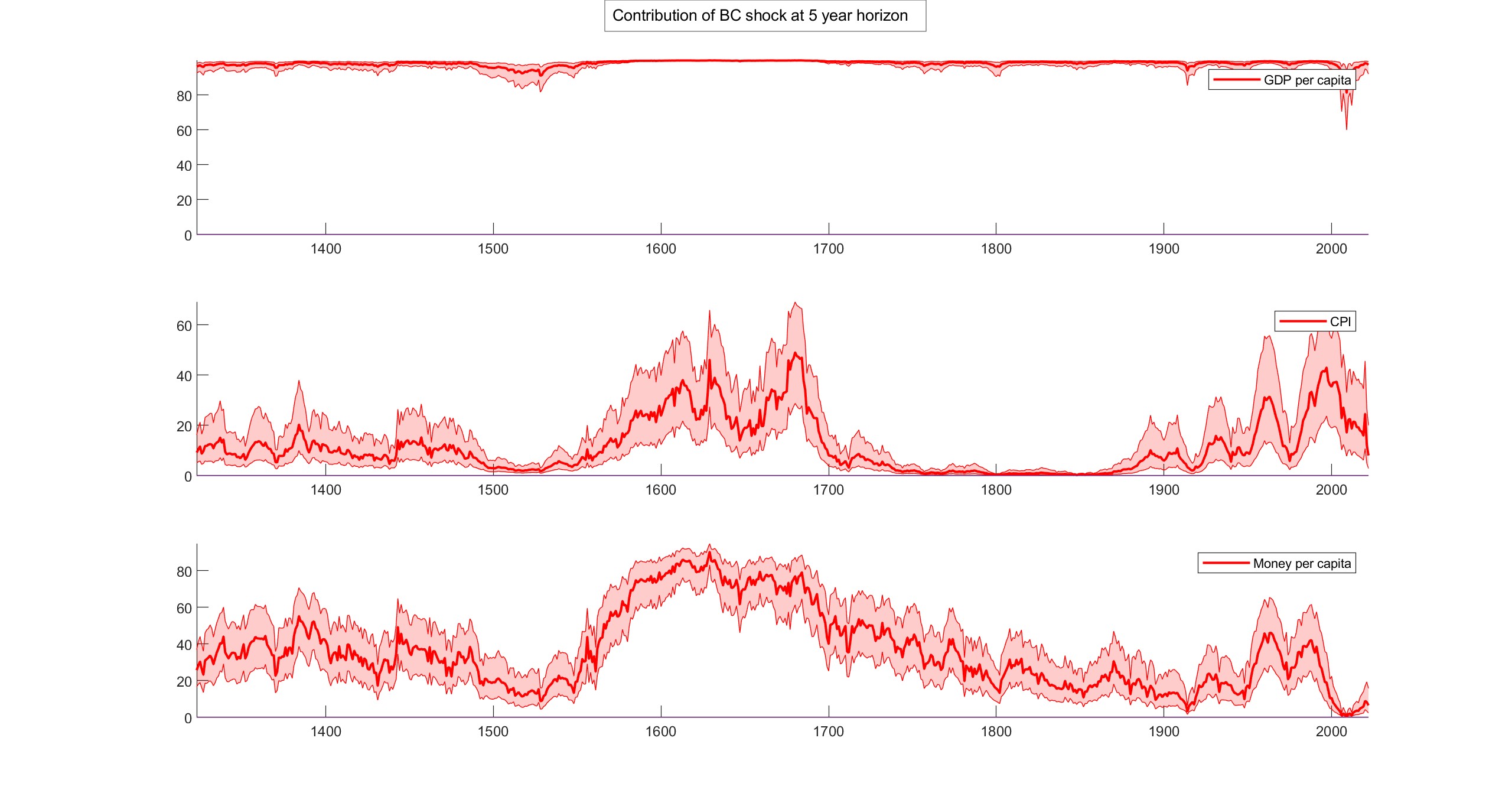} 
\par\end{centering}
}
\par\end{centering}
{\footnotesize Notes: The pink shaded areas represent the 68 percent
credible sets and the red lines depict the medians. }{\footnotesize\par}
\end{figure}

\begin{figure}[!tp]
\caption{Conditional Volatility\protect}
\label{fig:IRFs-1} 
\begin{centering}
\includegraphics[width=1\textwidth]{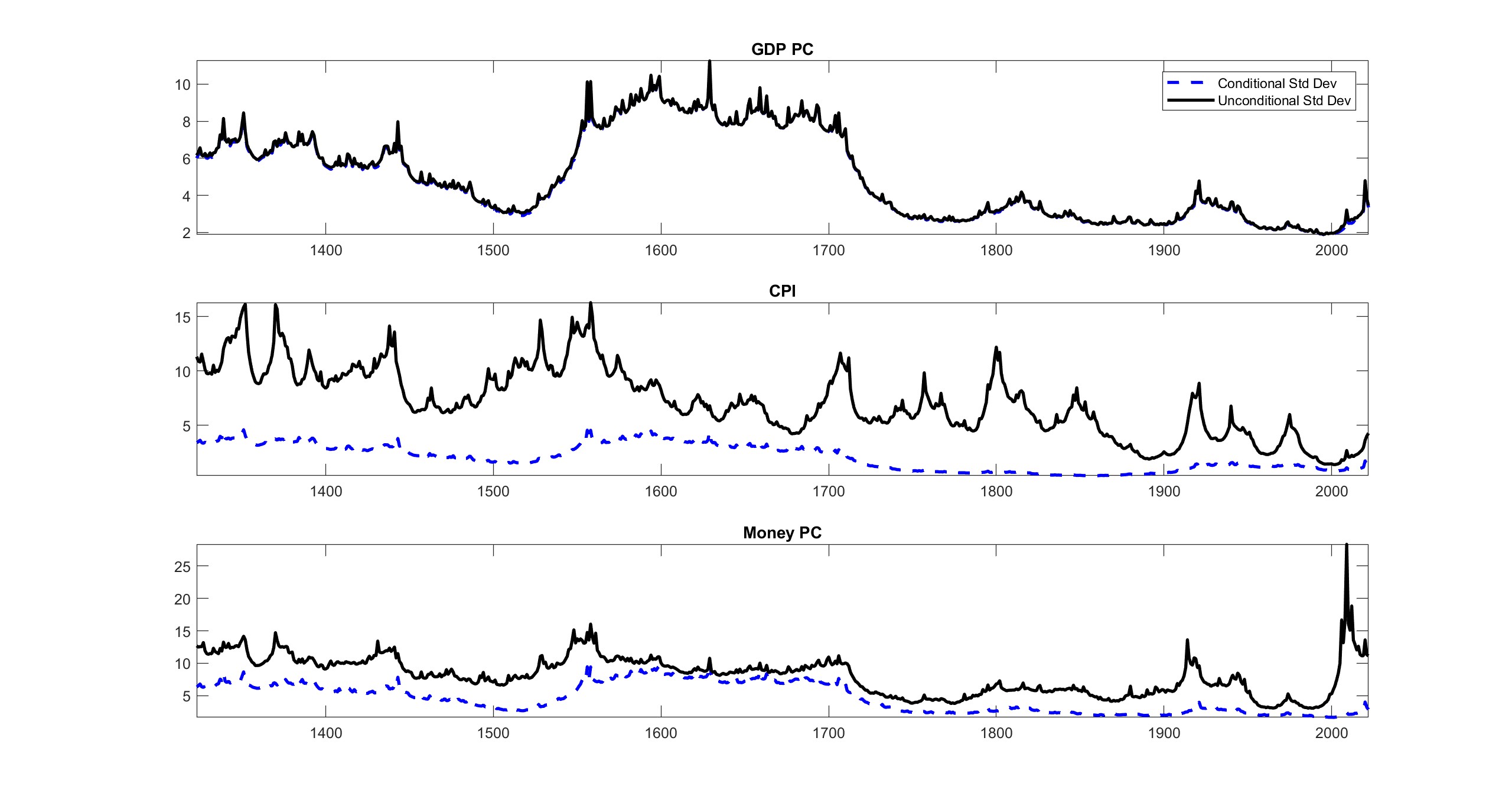} 
\par\end{centering}
{\footnotesize Notes: The dashed blue lines are the estimates of the
time-varying volatilities of each variable conditional on the business
cycle shock. The solid black lines are the estimates of the unconditional
volatilities of each of these variables. }{\footnotesize\par}
\end{figure}

The dynamic effects of business cycle shocks over time are vividly
illustrated in Figure \ref{fig:IRFs-3}, which presents the impulse
response functions (IRFs) and forecast error variance decompositions
(FEVDs) associated with these shocks. Figure \ref{fig:IRFs} shows
the IRFs cumulated over the 5-year horizon corresponding to a business
cycle shock that is normalised to have a unit standard deviation.
The top panel of Figure \ref{fig:IRFs-3} illustrates the responses
of GDP per capita, the middle panel shows the responses of CPI, and
the bottom panel depicts the responses of money stock per capita.
Moreover, the black lines in Figure \ref{fig:IRFs-1} represent the
estimated time-varying stochastic volatility, while the blue lines
capture the estimated volatility conditional on the identified business
cycle shock.

\paragraph{Supply Shocks as Drivers of Pre-Modern Business Cycles }

In earlier centuries, business cycle shocks exhibit a classic supply-side
pattern, where output declines while inflation rises (Figure \ref{fig:IRFs}).
This pattern aligns with the frequent occurrence of wars and famines
during this period.\footnote{Appendix Figure \ref{fig:Narrative-evidence} highlights significant
realisations of the business cycle shocks associated with well-documented
historical events characterized by supply-side dynamics.} While extreme events such as the Black Death (1348--50) resulted
in a larger contraction in population than in output---thereby increasing
output per capita---our estimates indicate that typical supply-side
shocks during this era tended to reduce output per capita while driving
up prices. Prior to 1500, agricultural production, particularly arable
farming, was highly volatile due to weather-related shocks, while
industrial production and commercial or monetary factors contributed
minimally to growth dynamics (\citealp{Broadberry2012wp}).

%\cleardoublepage

\paragraph{The Most Volatile Economic Period}

The 17th century stands out as the period with the strongest propagation
of business cycle shocks exhibiting supply-side characteristics. During
this time, a realization of the shock had the largest effects on GDP
and CPI, with changes of approximately 5\% and -4\%, respectively.
This aligns with the significant economic, social, and political upheaval
of the era, as reflected in the estimated stochastic volatility of
output, which peaks during this period (top panel of Figure \ref{fig:IRFs-1}).
The 17th century was marked by a global crisis characterized by frequent
wars, climate change, and widespread catastrophes (\citealp{parker2013}).\footnote{\citet{parker2013} argues that the \textquotedbl Little Ice Age\textquotedbl{}
brought prolonged cooling, erratic weather patterns, and repeated
harvest failures, leading to famines and economic instability. These
environmental stresses coincided with the English Civil War (1642--1651)
and widespread rebellions across Europe, which disrupted trade, devastated
agricultural output, and strained public finances. The interplay between
climate-induced scarcity and human-driven conflict created a \textquotedbl fatal
synergy\textquotedbl{} that destabilized the economy, reduced population
levels, and fuelled social unrest. In the UK, these dynamics were
further exacerbated by political upheaval, including the execution
of Charles I and the subsequent establishment of the Commonwealth.
This combination of natural disasters and human agency intensified
the magnitude and frequency of economic shocks, making the 17th century
one of the most volatile periods in British history.} Even in comparison to the 20th century, the business cycle volatility
of this period remains unparalleled.

The behaviour of inflation and money offers valuable insights into
the historical relationship between monetary drivers of inflation---or
the lack thereof---in earlier centuries. The negative conditional
comovement between inflation and money growth suggests that money
supply shocks, as posited by the quantity theory of money \citet{friedman1956quantity},
are unlikely to be the primary drivers of the identified business
cycle shocks. To contextualise this finding, historians have long
debated whether the inflow of precious metals from the Americas fuelled
the elevated inflation levels of the 16th century (\citet{hamilton1934american})
or whether inflationary pressures were instead driven by real factors,
such as demographic and resource constraints (\citet{ramsey1971price}).\footnote{See \citet{mccloskey1972price} for a review of the debate.}

Our empirical results do not definitively corroborate or refute either
hypothesis, as the forecast error variance in inflation (middle panel
of Figure \ref{fig:IRFs-2}) attributed to business cycle shocks remains
limited during most of the 16th century, when the price surge occurred.\footnote{See Figure 5.01 of \citet{broadberry2015british}.}Specifically,
business cycle shocks explain, at most, around 20\% of the forecast
error variance in inflation during the 16th century and approximately
40\% during the 17th century. However, the variance in inflation that
can be attributed to business cycle shocks appears inconsistent with
channels related to the quantity theory of money, given the negative
comovement between inflation and money growth induced by these shocks
during this period.

\paragraph{The Real Effects of Monetisation in the 16th and 17th Centuries}

Despite the limited ability of the business cycle shock to explain
inflation variation, it accounts for a substantial portion of the
variation in money during the 17th century. Beginning in the mid-16th
century, the forecast error variance in money rises from around 20\%
to nearly 80\% by 1600 and remains elevated throughout the 17th century.
This significant increase in the conditional comovement between the
business cycle and money aligns with the earlier inflow of precious
metals into the macroeconomy, lending empirical support to the argument
that ``monetisation greased the wheels of economic activity'' (\citealp{DobadoGonzalez2021}).
Empirical evidence also highlights the role of increased money supply
in stimulating economic growth (\citealp{Palma2018supply}).\footnote{Endogenous growth models (\citealp{romer1990JPE,aghion1992ec}) extended
with monetary features (\citealp{aghion2012,chu2014,gertler2016_technology,gil2010})
provide a theoretical rationale. } A proposed mechanism is that the greater availability of money enhanced
market participation and facilitated tax collection (\citealp{palma2018modernisation}).\footnote{Search-based theories of money (e.g. \citealp{shi1997}) could possibly
rationalise why increased money growth can increase agents' probability
of having a successful match.} Increased market participation, in turn, encouraged individuals to
work additional days and hours, thereby expanding the labor supply---a
phenomenon described as ``an industrious revolution'' (\citealp{deVries2008}).
Our findings provide empirical support for these arguments, particularly
in demonstrating the unprecedentedly large real effects associated
with money variation during the 17th century.

During the period 1700-1900, macroeconomic volatility fell (Figure
\ref{fig:IRFs-1}) and the business cycle shocks have a falling explanatory
power for money, suggesting that the real effects of monetisation
were fading, and the shock has an increasingly inflationary effect
during this period (middle panel of Figure \ref{fig:IRFs}). The latter
finding suggests an increasingly demand-like characteristic of the
business cycle shock, which coincides with the development of the
consumer society in Great Britain which was accelerated by the Industrial
Revolution (1760--1840). By the Victorian era (1837--1901), consumer
culture was firmly established, with growing emphasis on shopping
as a social activity and the acquisition of goods as a marker of identity
and status (\citealp{Perkin1969}).

\paragraph{Demand Shocks as Drivers of Modern Business Cycles }

After 1900, the explanatory power of the business cycle shock for
inflation rises markedly, consistent with an environment in which
inflation becomes more responsive to aggregate demand conditions.
Notably, this shift aligns with the broader transition to modern monetary
and fiscal frameworks as economies moved away from commodity-based
monetary systems, such as the gold standard, toward fiat currency
regimes. This transition granted governments greater flexibility to
manage aggregate demand through monetary and fiscal policies, leading
to inflation dynamics that were increasingly tied to economic cycles.
The adoption of these modern frameworks, particularly after World
War II, facilitated more active policy interventions, making inflation
more sensitive to aggregate demand shocks. This evolution is consistent
with historical evidence highlighting the prevalence of inflation
crises in the 20th century, driven by fiscal and monetary expansions
aimed at stabilizing growth and employment (\citealp{rogoffcarmern2008}).
Notable recent papers, such as (\citealt{Smets2007,Christiano2014}),
highlight the importance of demand-type shocks in driving business
cycle fluctuations. Our results show that the importance of demand
shocks as drivers of business cycles is only limited to the twentieth
century onwards.

\paragraph{Money Growth and Financial Development in the 20th Century}

\begin{figure}[bp!]
\caption{Inflationary effects and monetary developments}

\begin{centering}
\subfloat[Inflationary effects and money growth\label{fig:moneys2}]{\begin{centering}
\includegraphics[width=0.8\textwidth]{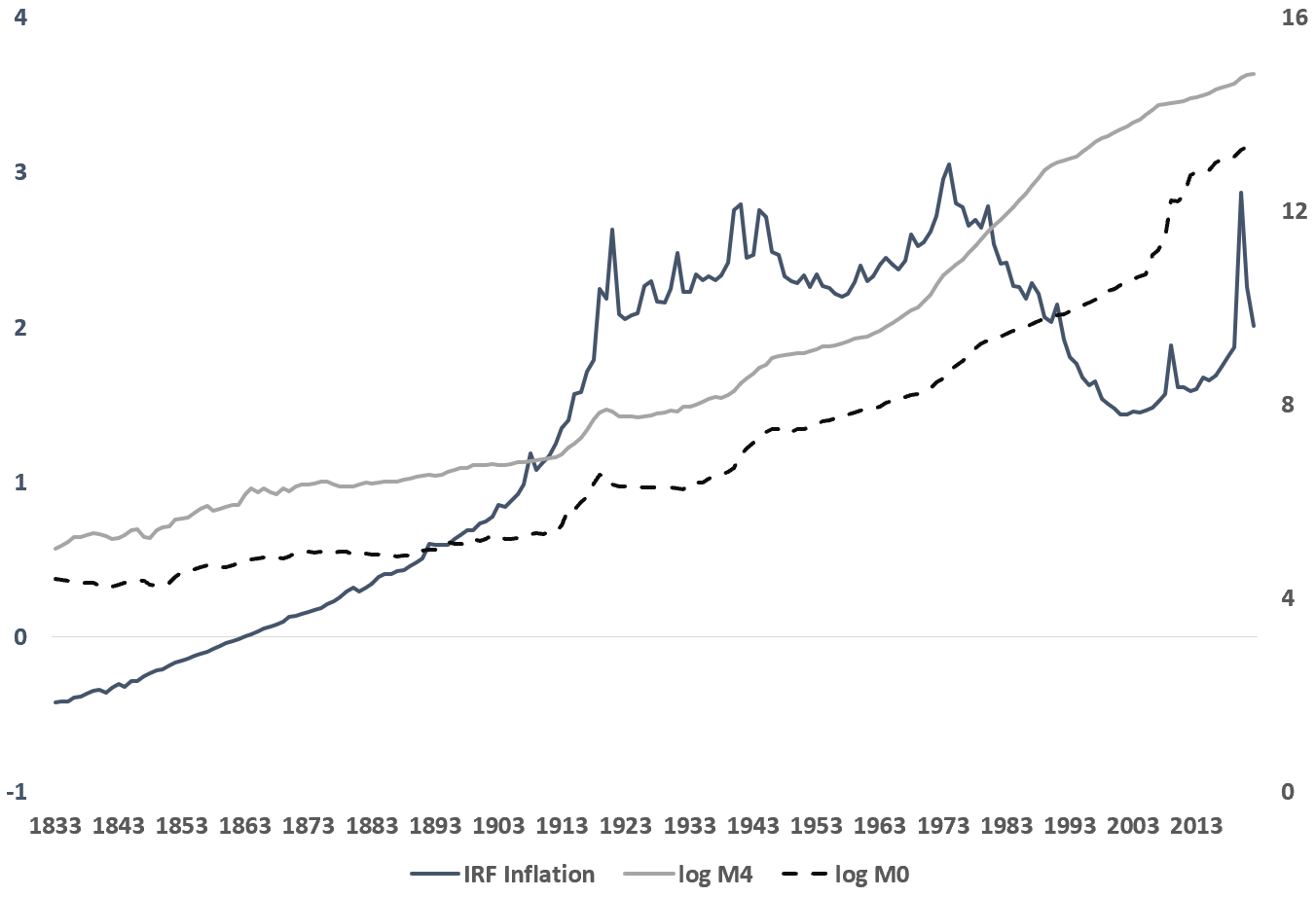} 
\par\end{centering}
}
\par\end{centering}
\begin{centering}
\subfloat[Inflationary effects and credit growth\label{fig:moneys}]{\begin{centering}
\includegraphics[width=0.8\textwidth]{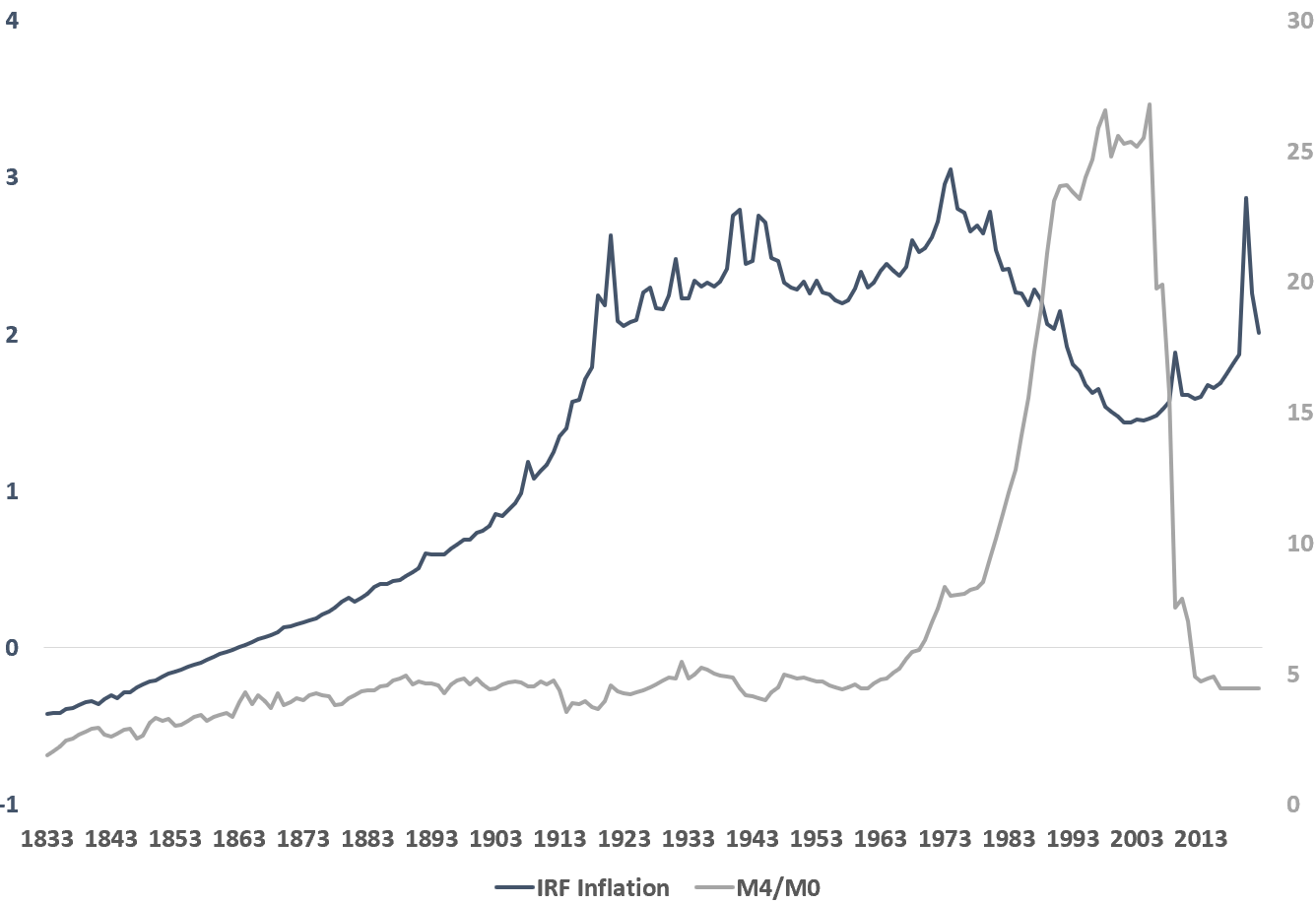} 
\par\end{centering}
}
\par\end{centering}
{\footnotesize Notes: The blue lines depict the the median IRFs of
inflation cumulated over the 5-year horizon and the other lines represent
the evolution of two measures of money (M4 and M0) as well as their
ratio. }{\footnotesize\par}
\end{figure}

The 20th century witnessed a profound transformation in the relationship
between money and macroeconomic dynamics. Figures~\ref{fig:moneys2}
and \ref{fig:moneys} illustrate this changing relationship, highlighting
the rapid expansion of both narrow money (M0) and broad money (M4)
starting in the early twentieth century. This expansion coincides
with the emergence of inflationary business cycle shocks, as shown
by the blue solid line in Figures~\ref{fig:moneys2}. This pattern
points to a structural break in the monetary regime and suggests that
monetary expansion---whether driven by loose monetary policy or endogenous
money demand---has increasingly shaped the inflation-output nexus
since 1900 (\citealp{rogoff2013}). 

To differentiate between monetary looseness and broader financial
deepening, we use the M4/M0 ratio as a proxy for financial development.
As illustrated in Figure \ref{fig:moneys}, this ratio shows a steady
increase after 1850, with a sharp acceleration post-1970. Interestingly,
during this period, the business cycle shock becomes less inflationary,
accompanied by a decline in inflation volatility. This observation
suggests that the expansion of the financial sector is not the primary
driver of inflationary pressure. Instead, the inflationary effects
of the business cycle shock in the twentieth century appear to be
more closely tied to the growth of base money (M0) rather than to
financial development itself.

To summarise, while the significant divergence between M0 and M4 occurred
more recently, the M4/M0 ratio rose from 1.9 in 1833 to 4.8 by 1900---a
period during which the inflation response shifted from negative to
positive. This suggests that the expansion of the money supply, rather
than financial development, is the more likely driver of the inflationary
nature of business cycle shocks in the twentieth century. Although
monetary expansion may result from loose monetary policy, shifts in
aggregate demand could also stimulate money growth through changes
in money demand. To disentangle these effects, we employ a four-variable
VAR model with interest rates and sign restrictions, as discussed
in the next section.

\paragraph{Decomposing Aggregate Demand Shocks}

\begin{figure}[!tp]
\caption{Decomposing aggregate demand: forecast error variance decomposition }
\label{fig:FEVD:-aggregate-demand} 
\begin{centering}
\subfloat[Pure aggregate demand shocks]{\begin{centering}
\includegraphics[width=1\textwidth]{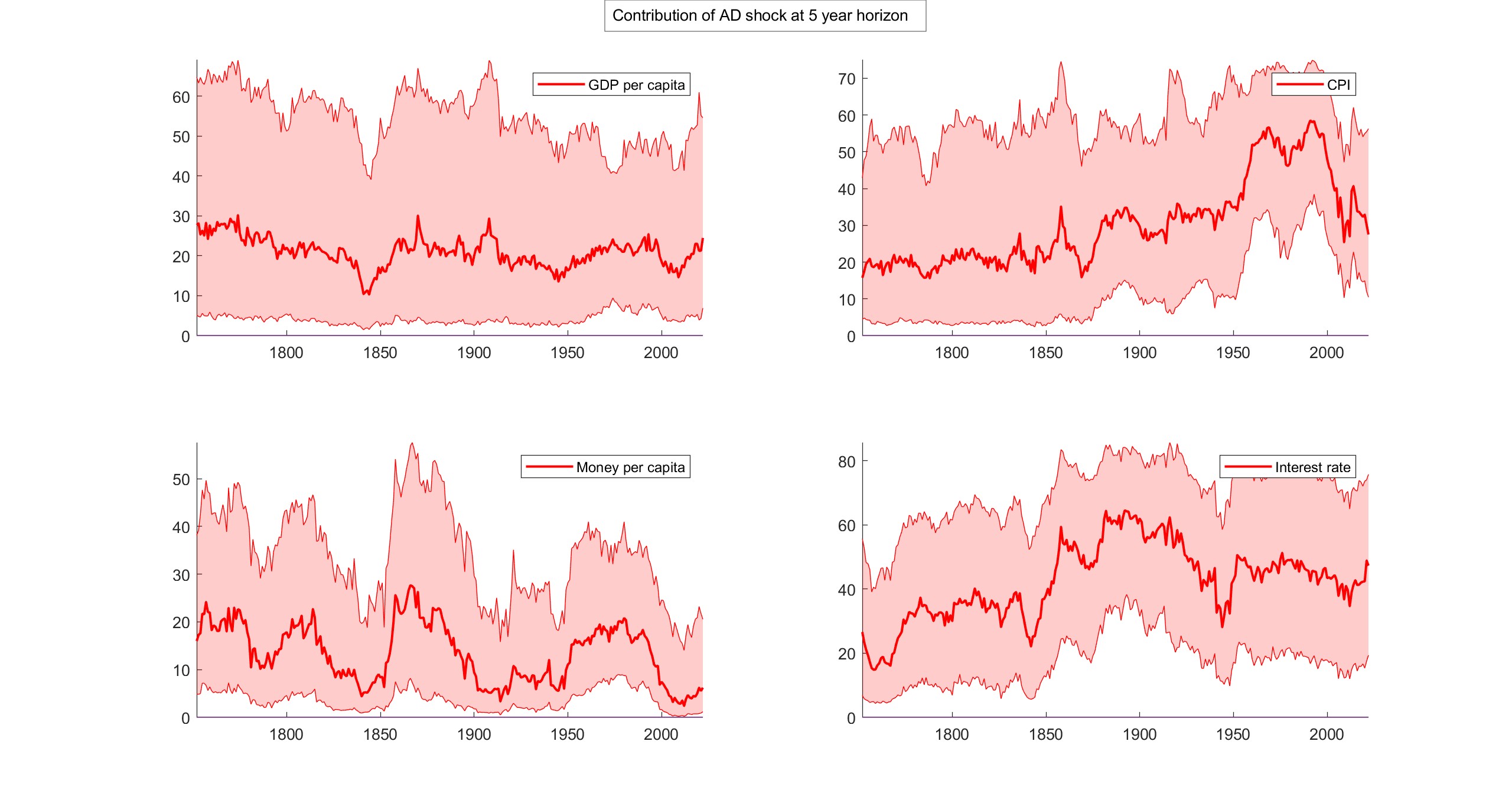} 
\par\end{centering}
}
\par\end{centering}
\begin{centering}
\subfloat[Money supply shocks]{\begin{centering}
\includegraphics[width=1\textwidth]{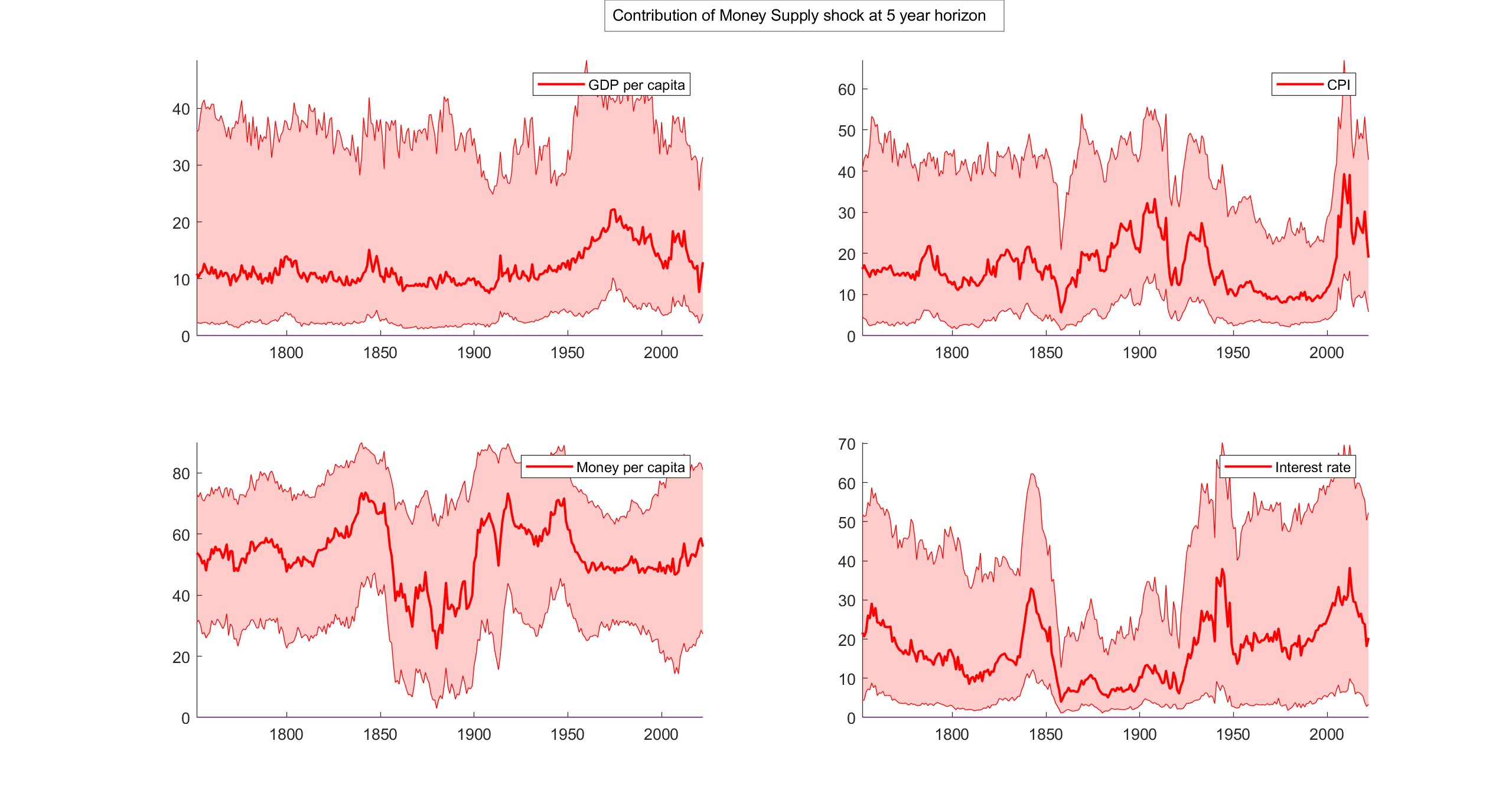} 
\par\end{centering}
}
\par\end{centering}
{\footnotesize Notes: The pink shaded areas represent the 68 percent
credible sets and the red lines depict the medians. }{\footnotesize\par}
\end{figure}

To separate monetary supply shocks from 'pure' aggregate demand shocks
(e.g., shifts in household preferences as described in \citet{Smets2007}),
we estimate a four-variable VAR that incorporates the nominal interest
rate into our baseline empirical model. While the inclusion of interest
rates limits the analysis to the post-1700 period, this extension
provides valuable insights. Figure \ref{fig:FEVD:-aggregate-demand}
presents the forecast error variance decompositions for the four variables
in the VAR with sign restrictions.

Despite the wide probability bands, both shocks appear to significantly
influence business cycle fluctuations in modern times. The median
estimates indicate that pure aggregate demand shocks account for a
larger share (20--30\%) of output variation compared to money supply
shocks (approximately 10--20\%). Similarly, pure demand shocks play
a more prominent role in explaining price level variation than money
supply shocks. In contrast, money supply shocks explain nearly 60\%
of the variation in money, aligning with the emergence of modern monetary
policy. Notably, the real effects of money variation appear more subdued
in modern times compared to the pronounced impacts observed in the
17th century, as discussed earlier.

\section{Robustness and extensions}

The conclusions reached in the previous section are robust to changes
in the information set and even to a change of model. In particular,
we replace M0 per capita with a broader measure of money: M4 per capita.\footnote{Among other holdings, M4 includes private sector's holdings of commercial
paper, bonds, Floating Rate Notes and other instruments of up to and
including five years' original maturity issued by UK Monetary Financial
Institutions.} Appendix Figure \ref{fig:irf_m4} shows that IRFs under this specification
are very similar to the baseline results, with the IRF of M4 shifting
slightly upward, especially over the last hundred years. Next, we
re-estimate the baseline regression without dividing GDP and M0 by
population. Appendix Figure \ref{fig:irfs_npc} shows that, once more,
IRFs remain very similar to the baseline results.

We also estimate a Dynamic Factor Model (DFM), which allows us to
expand the information set and address measurement error in an alternative
way. A short description of the model and results is presented in
Section \ref{dfm}. Once more, IRFs are similar to baseline results.
Moreover, since our DFM allows an unbalanced panel, we can include
\citet{Schmelzing2020boe}'s measure of long-term real interest rate
and estimate the natural rate of interest following \citet{lubik2015calculating}.
The resulting r-star (shown in Figure \ref{fig:rstar}) exhibits a
declining trend throughout the early centuries of the sample, followed
by a long period of relative stability. This stability was disrupted
in the late nineteenth century, with the natural rate peaking at 11.5\%
in 1974.

\section{Conclusion}

This paper documents how the nature and propagation of business cycle
shocks in the UK have evolved over many centuries. We show that pre-modern
cycles were dominated by supply-side shocks with stagflationary effects -- often linked to climate shocks, demographic events and wars --
while modern business cycles are predominantly driven by demand shocks. These findings highlight that business cycle dynamics are not time-invariant but are instead shaped by the prevailing
institutional and monetary environment, underscoring the importance
of historical perspective in understanding modern macroeconomic fluctuations. Monetisation, for instance, boosted real activity in earlier centuries but became increasingly inflationary as the economy transitioned to a modern monetary system.

We also show that output volatility
has declined over time, despite peaking in the seventeenth century, suggesting a centuries-long Great ``Moderation'', as economic development and stabilising institutions gradually reduced the amplitude of business cycle fluctuations.  This historical perspective reveals that even events  such as COVID-19 appear less anomalous when viewed against centuries of data. In fact, the volatilities of output and inflation observed during and in the aftermath of COVID-19 are far smaller than the values estimated for the periods before 1700.

Although our analysis is historical, the results carry important policy implications. First, institutional arrangements influence both the types of shocks economies face and the way these shocks propagate. Second, supply-driven stagflationary shocks have historically been the dominant form of macroeconomic disturbance. As shocks related to climate, energy, demographics and immigration become more common, understanding how supply shocks affected the economy in the past can help policymakers better prepare for disturbances whose dynamics may be similar to those of earlier centuries.

\newpage{} \bibliographystyle{chicago}
\bibliography{ref_new,big_semih}
 %Appendix
\newpage{}

\section*{Appendix}

\addcontentsline{toc}{section}{Appendices} 
\global\long\def\thesubsection{\Alph{subsection}}%

\setcounter{figure}{0} \setcounter{equation}{0} \setcounter{table}{0}

\counterwithin{figure}{subsection} \numberwithin{equation}{subsection}
\counterwithin{table}{subsection}

\subsection{Data}
\begin{itemize}
\item \underline{GDP per capita}: the measure of activity is the real GDP
(at market prices) per capita. Estimates for 1271-1700 period are
only available for England, so we use the estimated real GDP of England
throughout the entire sample. For the period where both UK and England
GDP series are available -- 1700 onwards -- the correlation between
their growth is very strong: 99\%. 
\item \underline{Inflation}: the measure of inflation is the CPI variation.
For 1271-1869, the series measures the cost of living in England only.
For 1870 onwards, it measures inflation in the entire UK. 
\item \underline{Broad Money}: the measure of broad money is M4x (M4 excluding
intermediate OFCs) from 1997 on; from 1996 to 1963 the variation in
M4 is used to backward extrapolate it; from 1962 to 1880, the variation
of a synthetic M4 (M3 + building society shares + building society
deposits - building societies holdings of cash) is used to construct
the series. Then the variation of M3 is applied to construct the series
until 1870; and, finally, from 1869 to 1271, the variation of an M3
estimate is used. 
\item \underline{Money}: from 1833, the measure of money is the result
of the merging of three different series of M0. From 1832 to 1271,
the variation of an M3 estimate is used to backward extrapolate this
series. Thus, for the first centuries of the sample our measures of
money and broad money exhibit the same behaviour. 
\end{itemize}
\begin{figure}[H]
\caption{Data\protect}
\label{fig:Narrative-evidence} 
\begin{centering}
\includegraphics[width=1\textwidth]{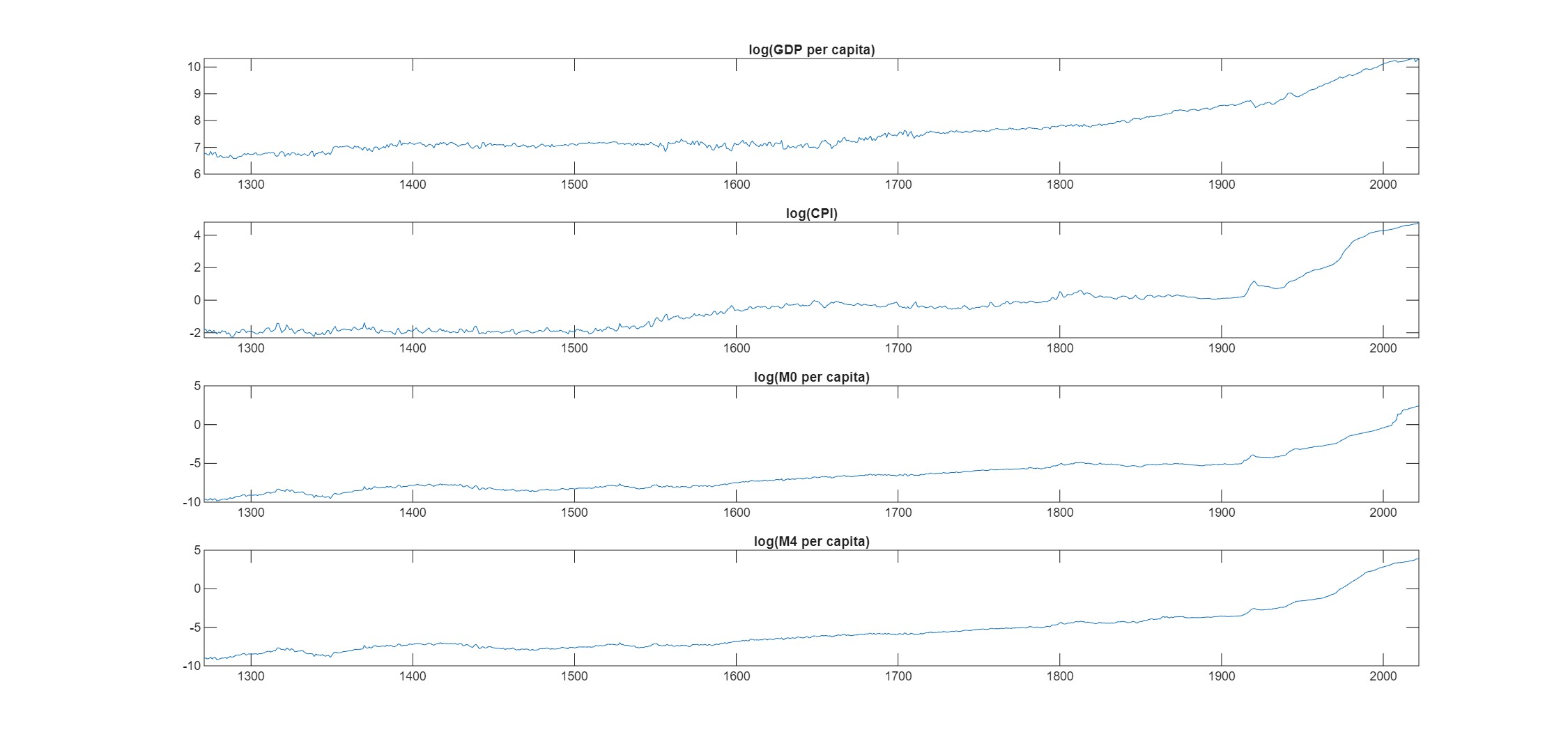} 
\par\end{centering}
{\footnotesize Notes: The blue lines depicts the log of the series
used in the estimation.}{\footnotesize\par}

\end{figure}

\subsection{Prior distributions and starting values}

The initial conditions for the VAR coefficients $\phi_{0}$ are obtained
via an OLS estimate of a fixed coefficient VAR using the first 50
observations of the sample period.

Let $\hat{v}^{ols}$ denote the OLS estimate of the VAR covariance
matrix estimated on the pre-sample data. For identification, the initial
value of the stochastic volatility $\sigma_{0}^{2}$ is fixed to the
diagonal elements of $\hat{v}^{ols}.$

The prior for the off-diagonal elements $A_{t}$ is \ $A_{0}\thicksim N\left(\hat{a}^{ols},V\left(\hat{a}^{ols}\right)\right)$
where $\hat{a}^{ols}$ are the off-diagonal elements of $\hat{v}^{ols}$,
with each row scaled by the corresponding element on the diagonal.
$V\left(\hat{a}^{ols}\right)$ is assumed to be diagonal with the
elements set equal to 10 times the absolute value of the corresponding
element of $\hat{a}^{ols}.$ The prior on $Q$ is assumed to be inverse
Wishart $Q_{0}\thicksim IW\left(\bar{Q}_{0},T_{0}\right)$ where $\bar{Q}_{0}$
is assumed to be $var(\hat{\phi}^{OLS})\times10^{-4}$ and $T_{0}$
is the length of the sample used to for calibration. The results are
not sensitive to this prior. We obtain similar results for larger
values of the scaling parameter. The prior distribution for the blocks
of $S$ is inverse Wishart: $S_{i,0}\thicksim IW(\bar{S}_{i},K_{i})$
where $i$ indexes the blocks of $S.$ $\bar{S}_{i}$ is calibrated
using $\hat{a}^{ols}$. Specifically, $\bar{S}_{i}$ is a diagonal
matrix with the relevant elements of $\hat{a}^{ols}$ multiplied by
$10^{-3}.$ Following \citet{Cogley_Sargent_2005} we postulate an
inverse-gamma distribution for the elements of $G$, $\sigma_{i}^{2}\sim IG\left(\frac{10^{-4}}{2},\frac{1}{2}\right)$.

Following \citet{koop03}, the prior for $\lambda_{k,i}=1/\varpi_{k,i}$
is gamma 
\begin{eqnarray*}
 &  & p\left(\lambda_{k}\right)\symbol{126}\Gamma\left(1,v_{\lambda,k}\right)\\
 &  & p\left(v_{\lambda,k}\right)\symbol{126}\Gamma\left(v_{0},2\right)
\end{eqnarray*}
where $v_{0}=20$ and $\Gamma\left(a,b\right)$ is the gamma density
with mean $a$ and df $b$.

\subsection{Simulating the posterior distribution}

\label{Gibbs}

We use a Gibbs sampling algorithm to sample from the posterior distribution.
The details of each conditional distribution are provided below.

\subsubsection{Time-varying VAR coefficients}

Conditional on the time-varying volatilities and contemporaneous coefficients,
the model is a SUR system with time-varying parameters. The \citet{ckohn2005}
algorithm is used to sample from the conditional posterior of $\phi_{t}.$
The distribution of the time-varying VAR coefficients $\phi_{t}$
conditional on all other parameters and hyperparameters is linear
and Gaussian: $\phi_{t}\backslash Z_{t},\Xi\thicksim N\left(\phi_{T\backslash T},P_{T\backslash T}\right)$
and $\phi_{t}\backslash\phi_{t+1,}Z_{t},\Xi\thicksim N\left(\phi_{t\backslash t+1,\phi_{t+1}},P_{t\backslash t+1,\phi_{t+1}}\right)$
where $t=T-1,..1,$ $\Xi$ denotes a vector that holds all the other
VAR parameters and $\phi_{T\backslash T}=E\left(\phi_{T}\backslash Z_{t},\Xi\right),P_{T\backslash T}=Cov\left(\phi_{T}\backslash Z_{t},\Xi\right),\phi_{t\backslash t+1,\phi_{t+1}}=E\left(\phi_{t}\backslash Z_{t},\Xi,\phi_{t+1}\right)$
and $P_{t\backslash t+1,F_{t+1}}=Cov\left(\phi_{t}\backslash Z_{t},\Xi,\phi_{t+1}\right).$
As shown by \citet{ckohn2005} the simulation proceeds as follows.
First we use the Kalman filter to draw $\phi_{T\backslash T}$ and
$P_{T\backslash T}$ and then proceed backwards in time using $\phi_{t|t+1}=\phi_{t|t}+P_{t|t}P_{t+1|t}^{-1}\left(\phi_{t+1}-\phi_{t}\right)$
and $\phi_{t|t+1}=\phi_{t|t}-P_{t|t}P_{t+1|t}^{-1}P_{t|t}.$

\subsubsection{Elements of $H_{t}$}

Following \citet{Cogley_Sargent_2005}, the diagonal elements of the
VAR covariance matrix are sampled using the Metropolis Hastings algorithm
in \citet{jpolrossi04}. Given a draw for $\phi_{t}$ the VAR model
can be written as $A_{t}^{\prime}\left(\tilde{Z}_{t}\right)=u_{t}$.
where $\tilde{Z}_{t}=Z_{t}-\sum_{l=1}^{L}\phi_{l,t}Z_{t-l}=v_{t}$
and $VAR\left(u_{t}\right)=H_{t}.$ \citet{jpolrossi04} note that
conditional on other VAR parameters, the distribution $h_{it},i=1..N$
is given by $f\left(h_{it}\backslash h_{it-1},h_{it+1},u_{it}\right)=f\left(u_{it}\backslash h_{it}\right)\times f\left(h_{it}\backslash h_{it-1}\right)\times f\left(h_{it+1}\backslash h_{it}\right)=h_{it}^{-0.5}\exp\left(\frac{-u_{it}^{2}}{2h_{it}}\right)\times h_{it}^{-1}\exp\left(\frac{-\left(\ln h_{it}-\mu\right)^{2}}{2\sigma_{h_{i}}}\right)$
where $\mu$ and $\sigma_{h_{i}}$ denote the mean and the variance
of the log-normal density $h_{it}^{-1}\exp\left(\frac{-\left(\ln h_{it}-\mu\right)^{2}}{2\sigma_{h_{i}}}\right).$
\citet{jpolrossi04} suggest using $h_{it}^{-1}\exp\left(\frac{-\left(\ln h_{it}-\mu\right)^{2}}{2\sigma_{h_{i}}}\right)$
as the candidate generating density with the acceptance probability
defined as the ratio of the conditional likelihood $h_{it}^{-0.5}\exp\left(\frac{-u_{it}^{2}}{2h_{it}}\right)$
at the old and the new draw. This algorithm is applied at each period
in the sample.

\subsubsection{Elements of $A_{t}$}

Given a draw for $\phi_{t}$ the VAR model can be written as $A_{t}^{\prime}\left(\tilde{Z}_{t}\right)=u_{t}$
where $\tilde{Z}_{t}=Z_{t}-\sum_{l=1}^{L}\phi_{l,t}Z_{t-l}=v_{t}$
and $VAR\left(u_{t}\right)=H_{t}.$ This is a system of equations
with time-varying coefficients and given a block diagonal form for
$Var(\tau_{t})$ the standard methods for state space models described
in \citet{ckohn2005} can be applied.

\subsubsection{VAR hyperparameters}

Conditional on $Z_{t}$, $\phi_{l,t}$, $H_{t}$, and $A_{t}$, the
innovations to $\phi_{l,t}$, $H_{t}$, and $A_{t}$ are observable,
which allows us to\
draw the hyperparameters---, the elements of\ $Q$, $S$, and the
$\sigma_{i}^{2}$---, from their respective distributions.

\subsubsection{Elements of $\lambda_{t}$}

The conditional posterior distributions related to the t-distributed
shock structure of the model are described in \citet{koop03}. Note
that conditional on $\Phi_{t}$ and $A_{t}$, the orthogonalized residuals
$\ $can be obtained as $\tilde{\varepsilon}_{t}=A_{t}u_{t}$. The
conditional posterior distribution for $\lambda_{i,t}$ derived in
\citet{geweke1993} applies to each column of $\tilde{\varepsilon}_{t}$.
As shown in \citet{koop03}this posterior density is a gamma distribution
with mean $\left(v_{\lambda,i}+1\right)/\frac{1}{\sigma_{i,t}}\tilde{\varepsilon}_{i,t}^{2}+v_{\lambda,i}$
and degrees of freedom $v_{\lambda,i}+1$. Note that $\tilde{\varepsilon}_{i,t}$
is the i$th$ column of the matrix $\tilde{\varepsilon}_{t}$.

\subsubsection{Degrees of freedom $v_{\lambda,i}$}

The conditional distribution for the degree of freedom parameter is
non-standard and given by: 
\begin{equation}
G\left(v_{\lambda,i}\backslash\lambda_{i}\right)\propto\left(\frac{v_{\lambda,i}}{2}\right)^{\frac{Tv_{\lambda,n}}{2}}\Gamma\left(\frac{v_{\lambda,n}}{2}\right)^{-N}\exp\left(-\left(\frac{1}{v_{0}}+0.5\sum_{t=1}^{T}\left[\ln\left(\lambda_{t,n}^{-1}\right)+\lambda_{t,n}\right]\right)v_{\lambda,n}\right)\label{dof}
\end{equation}
As in \citet{geweke1993} we use the Random Walking Metropolis Hastings
Algorithm to draw from this conditional distribution. More specifically,
for each of the $n$ equations of the VAR, we draw $v_{\lambda,n}^{new}=v_{\lambda,n}^{old}+c^{1/2}\epsilon$
with $\epsilon\sim N(0,1)$. The draw is accepted with probability
$\frac{G\left(v_{\lambda,n}^{new}\backslash\lambda_{n}\right)}{G\left(v_{\lambda,n}^{old}\backslash\lambda_{n}\right)}$
with $c$ chosen to keep the acceptance rate between $20\%$ and $40\%$.

\subsection{The Gibbs sampling algorithm}

The Gibbs algorithm for the model cycles through the following eight
conditional posterior distributions: 
\begin{enumerate}
\item $G(\lambda_{i,t}\backslash\Psi)$ where $\Psi$ denotes the remaining
parameters of the model. 
\item $G\left(v_{\lambda,i}\backslash\lambda_{i,t}\right)$ 
\item $G\left(g_{i,t}\backslash\Psi\right)$ 
\item $G\left(\sigma_{i,t}^{2}\backslash\Psi\right)$ 
\item $G\left(A_{t}\backslash\Psi\right)$ 
\item $G\left(\Phi_{t}\backslash\Psi\right)$ 
\item $G\left(Q\backslash\Psi\right)$ 
\item $G\left(Q_{A}\backslash\Psi\right)$ 
\end{enumerate}

\subsection{Additional results}

\subsubsection{Matching the Business Cycle Shocks with Narrative Accounts}

This figure shows that during episodes such as the Elizabethan famine
(1594--1597), the English Civil War (1642--1651), the Napoleonic
Wars (1812--1814), the 1816 crop failure, and the Irish Famine (1845--1852),
the model consistently identifies large negative business cycle shocks.
These are characterized by declining output and rising prices, confirming
the supply-side nature of macroeconomic disruptions in this era (\citet{broadberry2015british}).
\begin{figure}[H]
\caption{Narrative evidence\protect}
\label{fig:Narrative-evidence}
\begin{centering}
\includegraphics[width=0.8\textwidth]{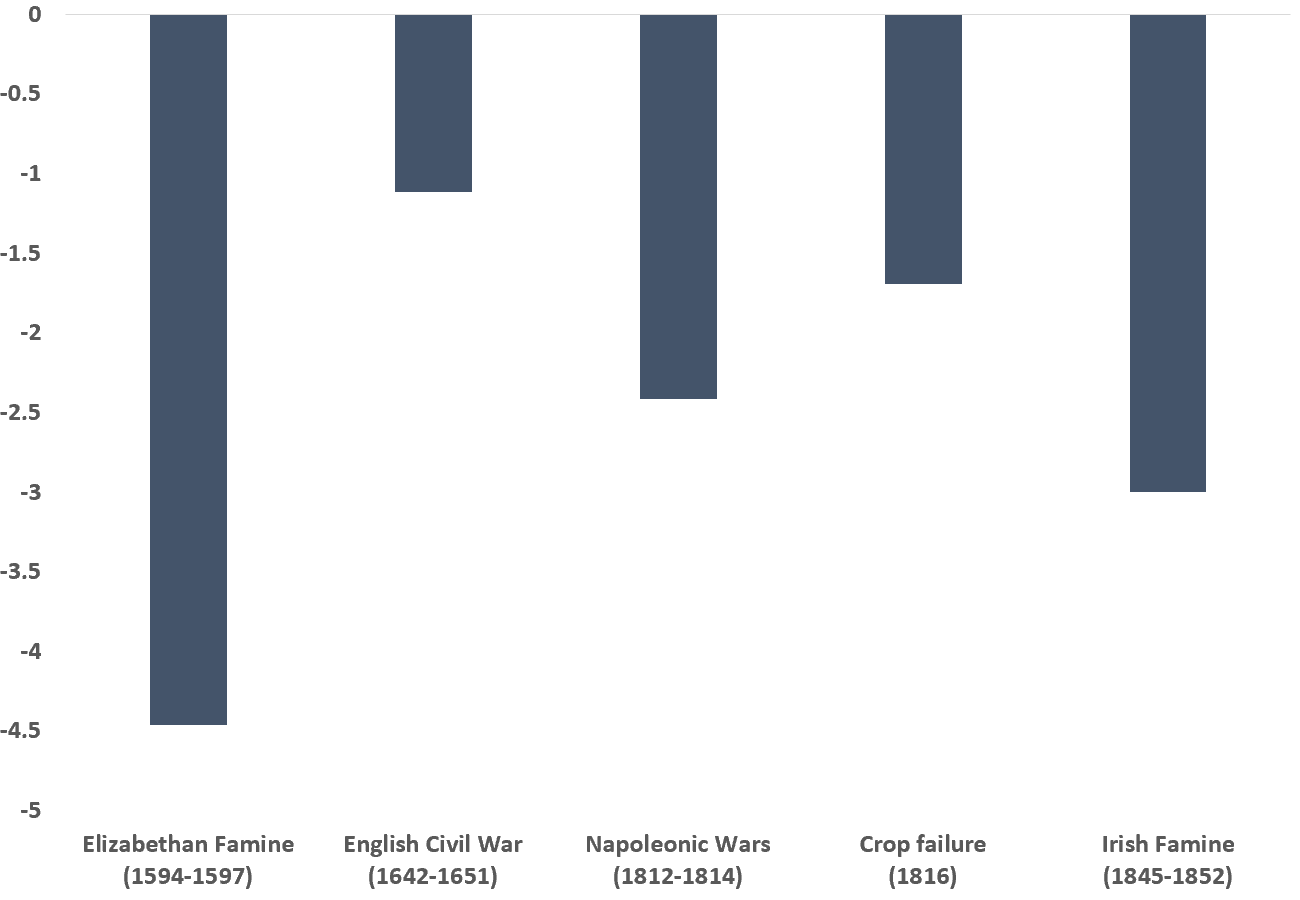} 
\par\end{centering}
{\footnotesize Notes: The bars show the main business cycle shocks
cumulated over the years for several famous episodes of British history.}{\footnotesize\par}

\end{figure}

\subsubsection{Predictability }

\begin{figure}[H]
\caption{Predictability\protect}
\label{fig:predictability} 
\begin{centering}
\includegraphics[width=1\textwidth]{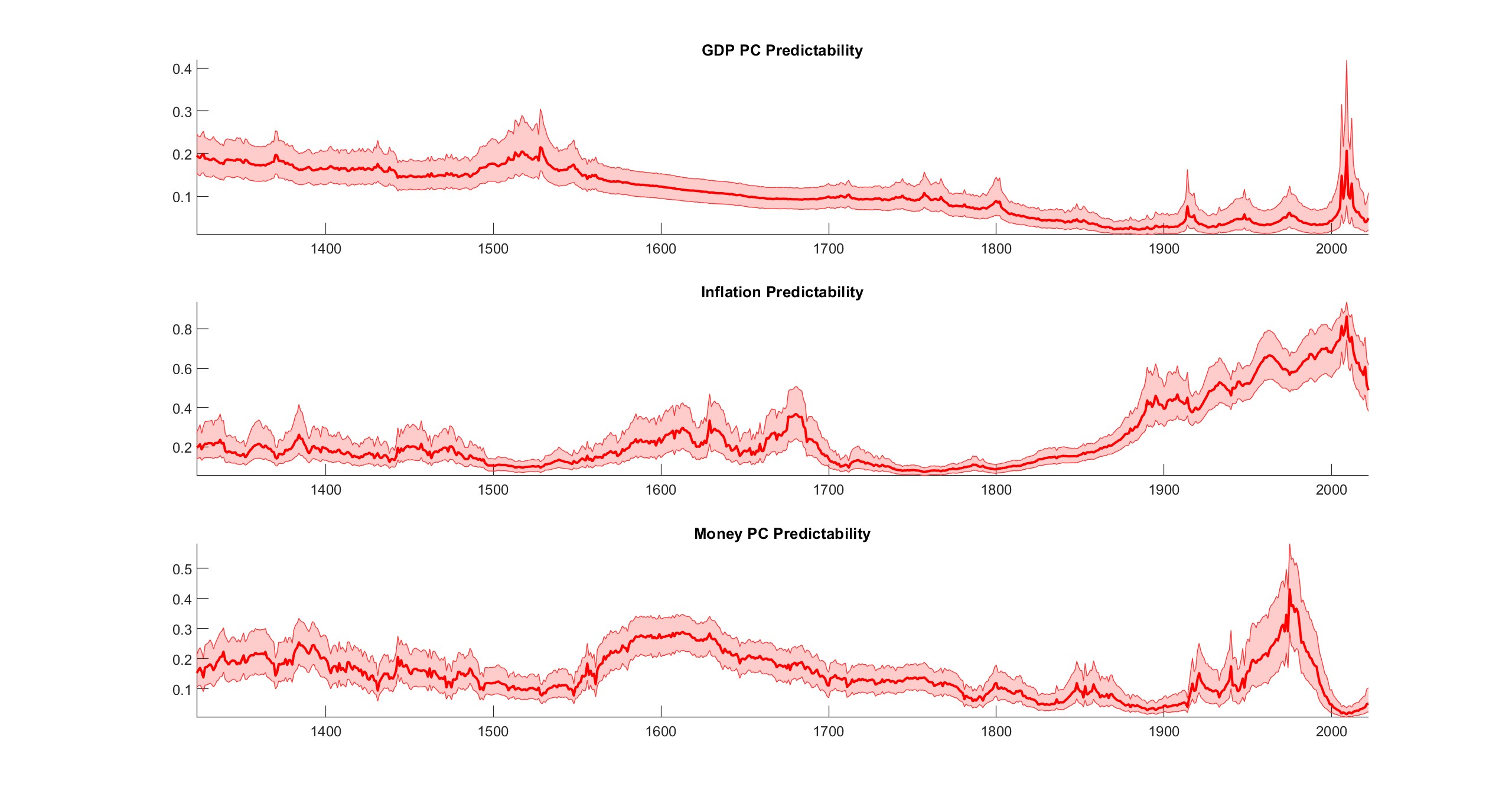} 
\par\end{centering}
{\footnotesize Notes: The pink shaded areas represent the 68 percent
credible sets and the red lines depict the medians. }{\footnotesize\par}
\end{figure}

We also examine the predictability of output, inflation, and money
using forecast $R^{2}$ (\citealp{cogley2010}) from the VAR model.
As shown in Figure \ref{fig:predictability}, the rise in inflation
predictability is a relatively recent phenomenon, consistent with
the findings of \citet{cogley2010} for post-war U.S. data. Similar
to \citet{nason2023}, we find that inflation predictability remained
low throughout earlier centuries. Predictability increases substantially
in the twentieth century. Output, on the other hand, shows a contrasting
trend: output predictability has actually declined since the 1300s
with the exception of the Great Financial Crisis. 

\subsubsection{Three-variable VAR with M4}

\begin{figure}[H]
\caption{Stochastic Volatility}

\begin{centering}
\includegraphics[width=1\textwidth]{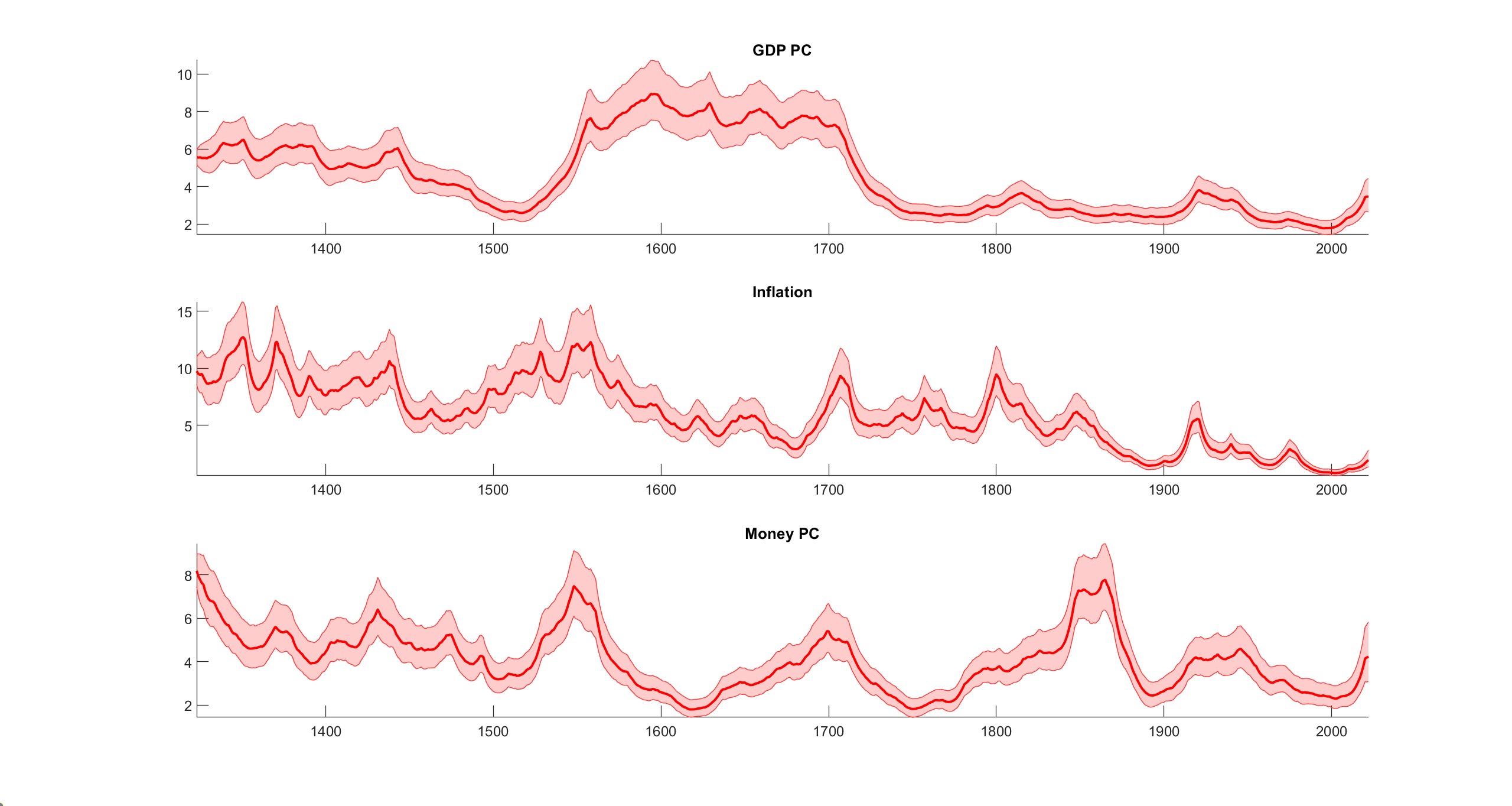} 
\par\end{centering}
{\footnotesize Notes: The pink shaded areas represent the 68 percent
credible sets and the red lines depict the medians. }{\footnotesize\par}
\end{figure}

\begin{figure}[H]
\caption{Impulse responses}

\begin{centering}
\includegraphics[width=1\textwidth]{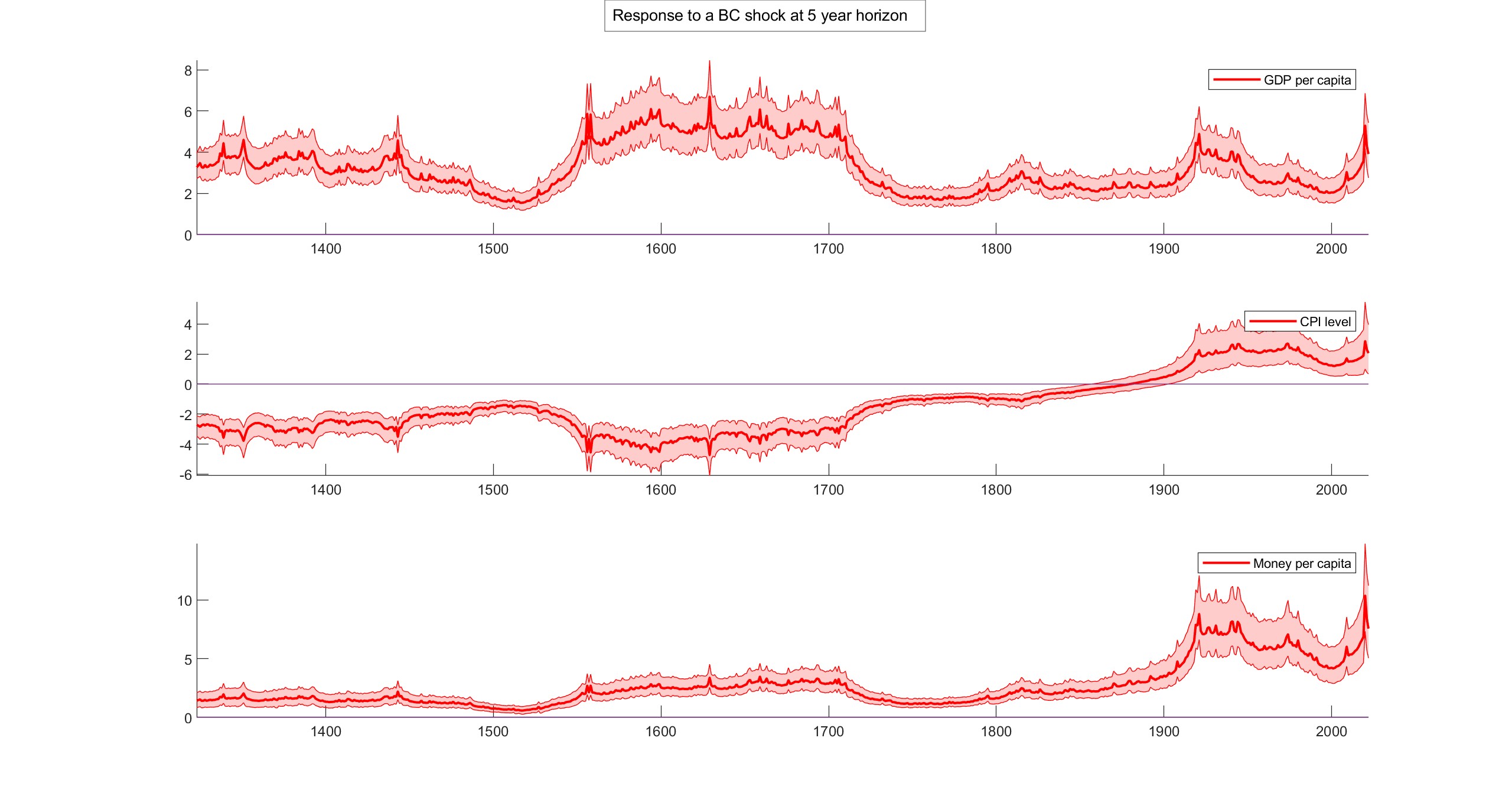} 
\par\end{centering}
\label{fig:irf_m4} {\footnotesize Notes: The pink shaded areas represent
the 68 percent credible sets for the IRFs and the red lines depict
the median IRFs using M4 as the measure of money. }{\footnotesize\par}
\end{figure}

\begin{figure}[H]
\caption{Conditional Volatility}

\begin{centering}
\includegraphics[width=1\textwidth]{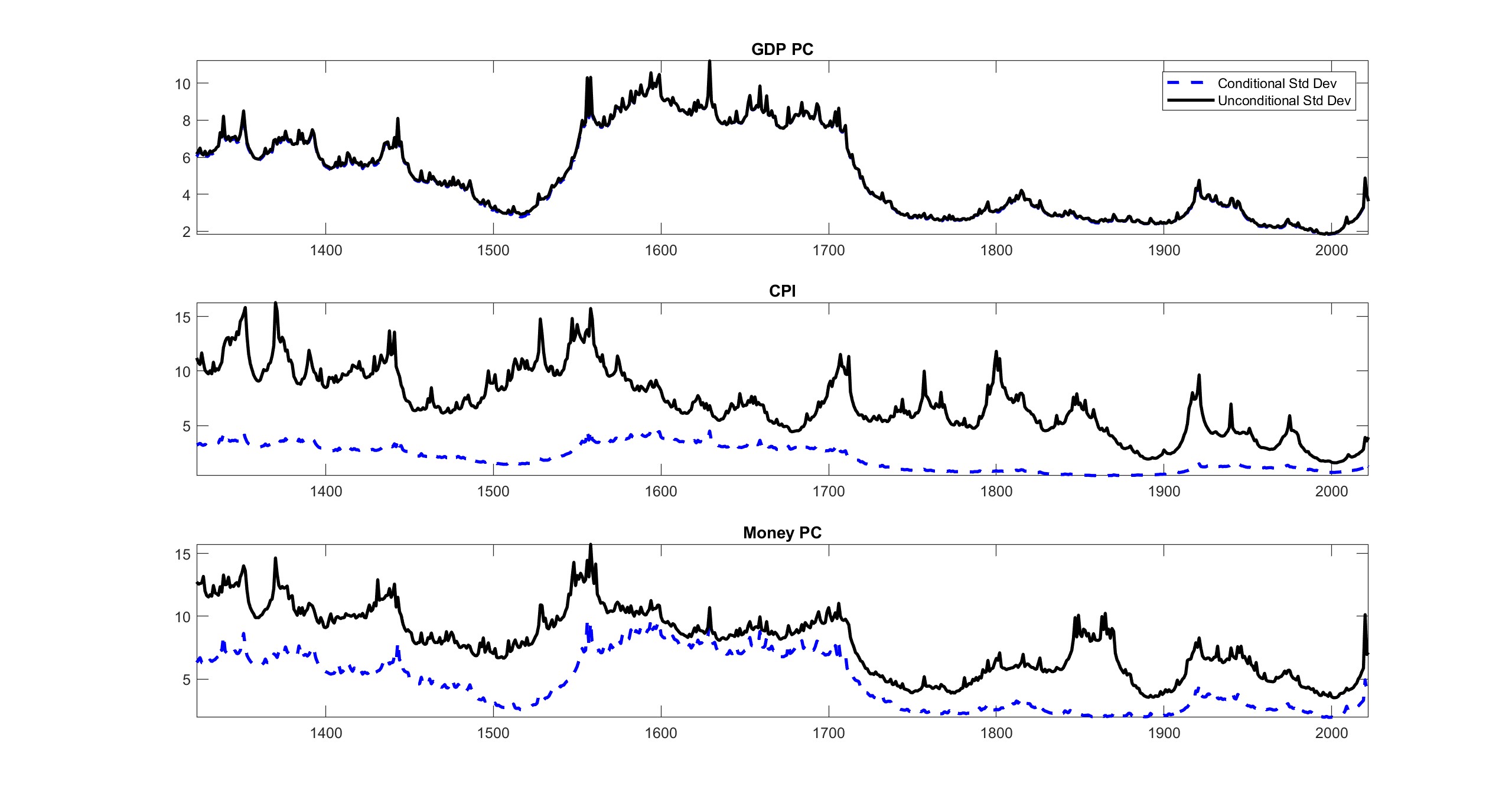} 
\par\end{centering}
{\footnotesize Notes: The dashed blue lines are the estimates of the
time-varying volatilities of each variable conditional on the business
cycle shock. The solid black lines are the estimates of the unconditional
volatilities of each of these variables. }{\footnotesize\par}
\end{figure}

\begin{figure}[H]
\caption{Predictability}

\begin{centering}
\includegraphics[width=1\textwidth]{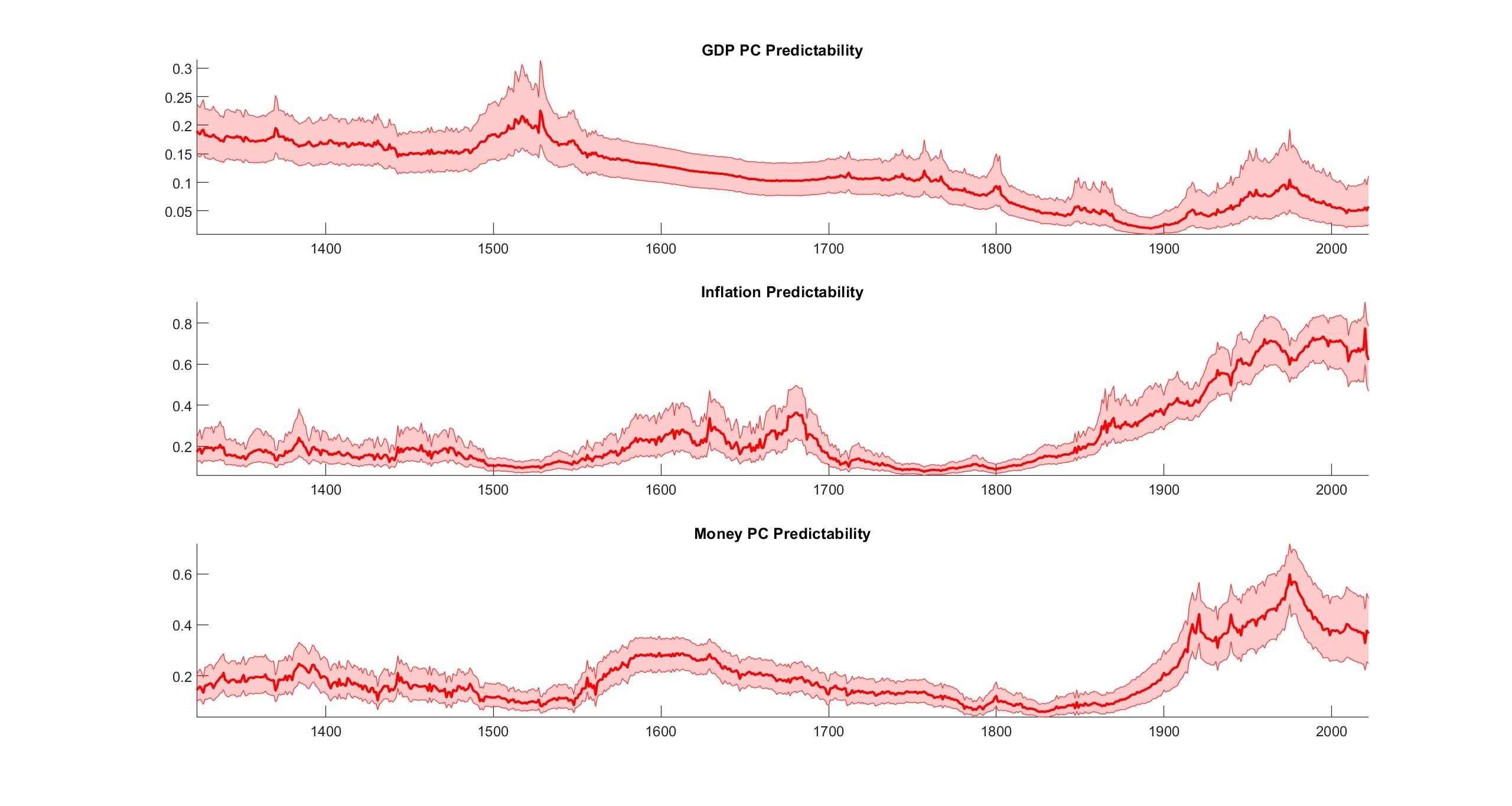} 
\par\end{centering}
{\footnotesize Notes: The pink shaded areas represent the 68 percent
credible sets and the red lines depict the medians. }{\footnotesize\par}
\end{figure}

\subsubsection{Three-variable VAR using aggregate (non-per-capita) variables}

\begin{figure}[H]
\caption{Stochastic Volatility}

\begin{centering}
\includegraphics[width=1\textwidth]{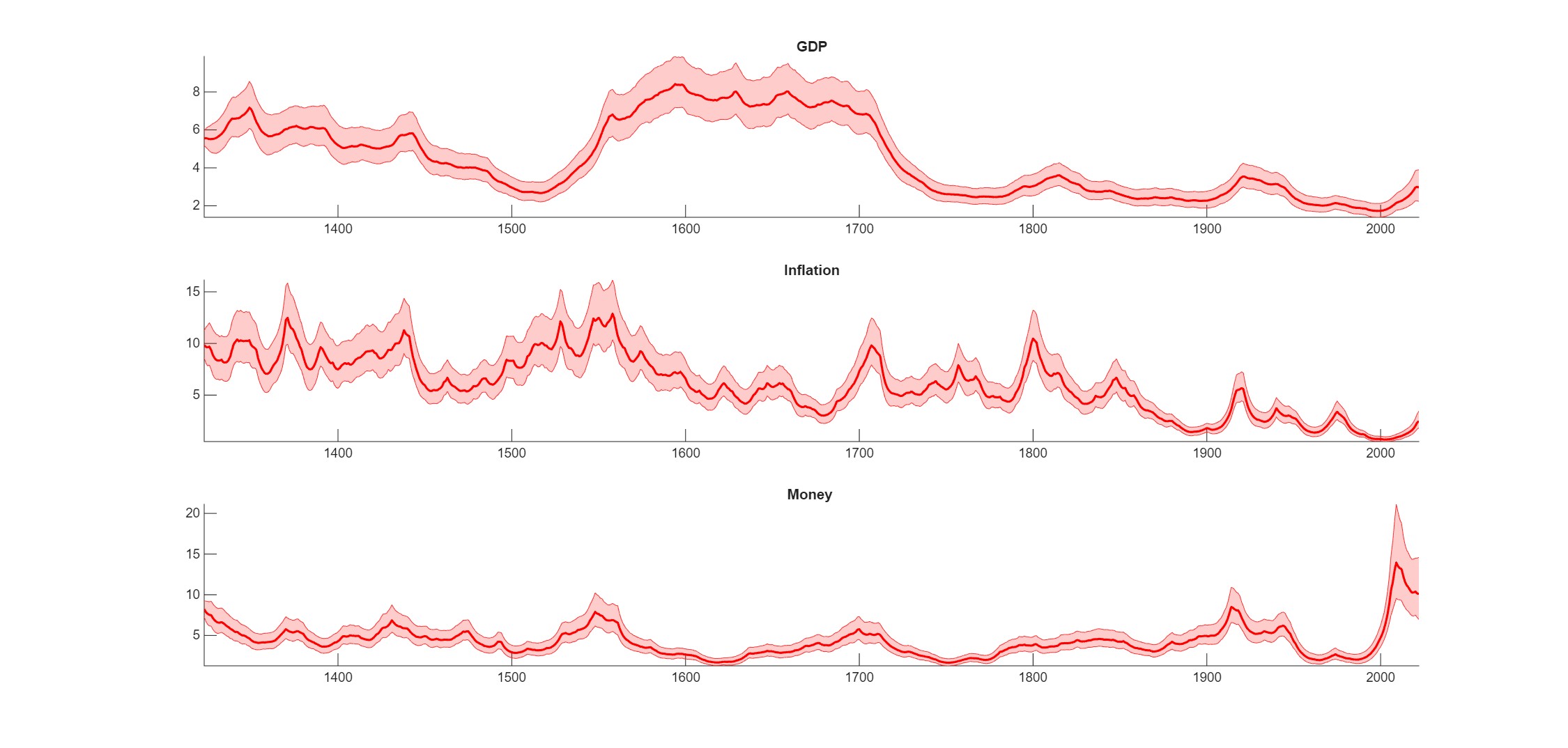} 
\par\end{centering}
{\footnotesize Notes: The pink shaded areas represent the 68 percent
credible sets and the red lines depict the medians. }{\footnotesize\par}
\end{figure}

\begin{figure}[H]
\caption{Impulse responses}

\begin{centering}
\includegraphics[width=1\textwidth]{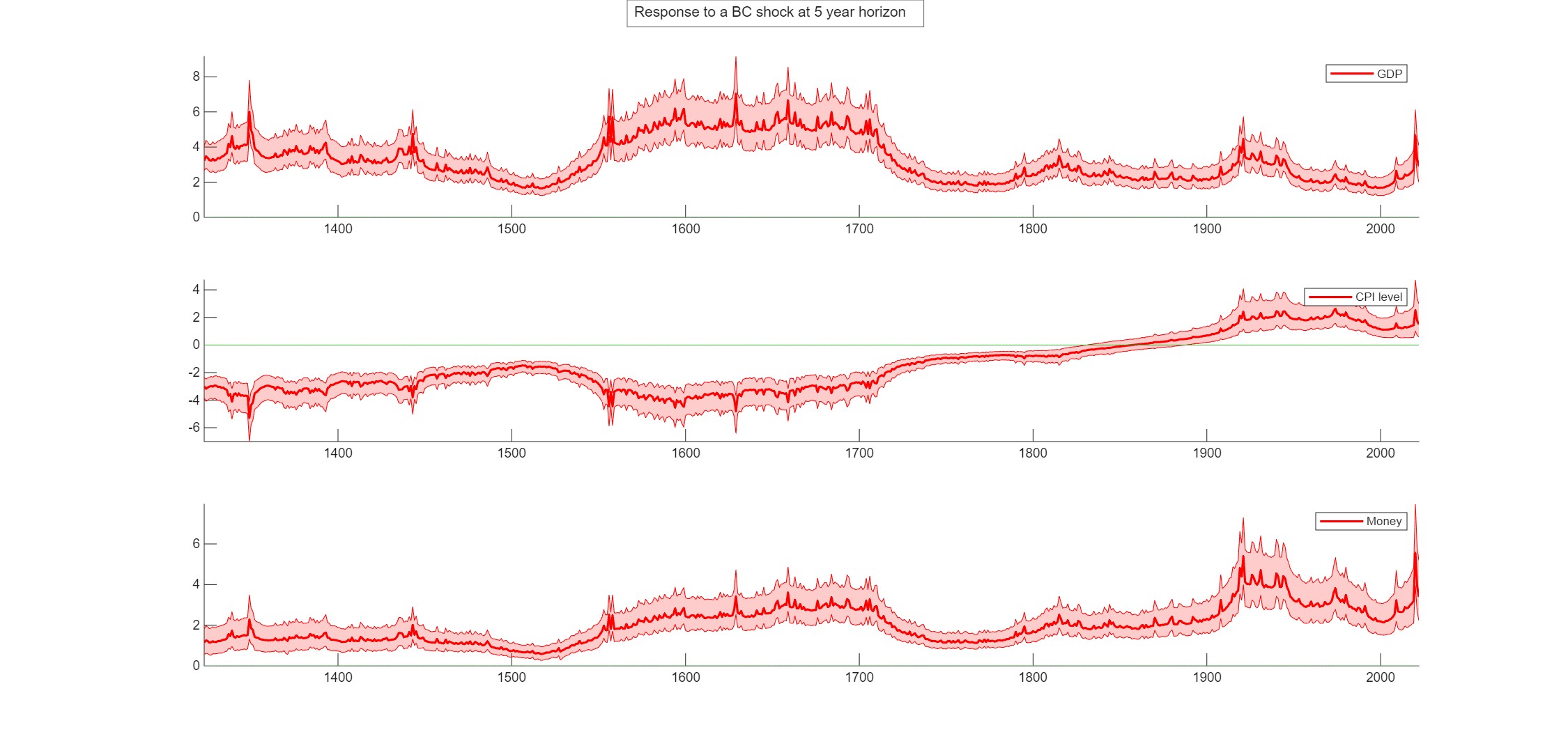} 
\par\end{centering}
\label{fig:irfs_npc} {\footnotesize Notes: The pink shaded areas represent
the 68 percent credible sets for the IRFs and the red lines depict
the median IRFs using aggregate (non-per-capita) variables. }{\footnotesize\par}
\end{figure}

\begin{figure}[H]
\caption{Predictability}

\begin{centering}
\includegraphics[width=1\textwidth]{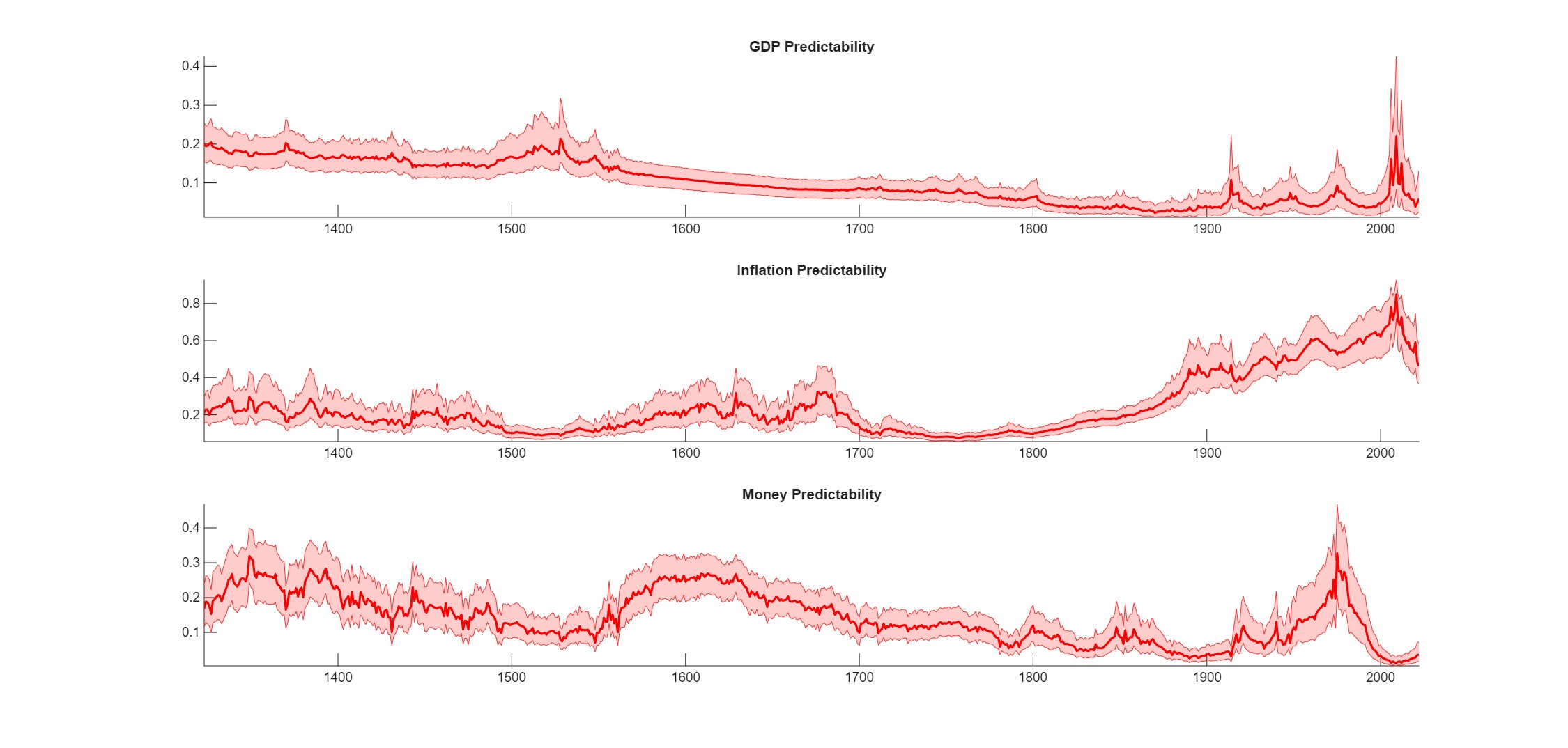} 
\par\end{centering}
{\footnotesize Notes: The pink shaded areas represent the 68 percent
credible sets and the red lines depict the medians. }{\footnotesize\par}
\end{figure}

\newpage{}

\subsubsection{Extra sign restriction plots}

\begin{figure}[H]
\caption{IRF: aggregate demand}

\begin{centering}
\includegraphics[width=0.95\textwidth]{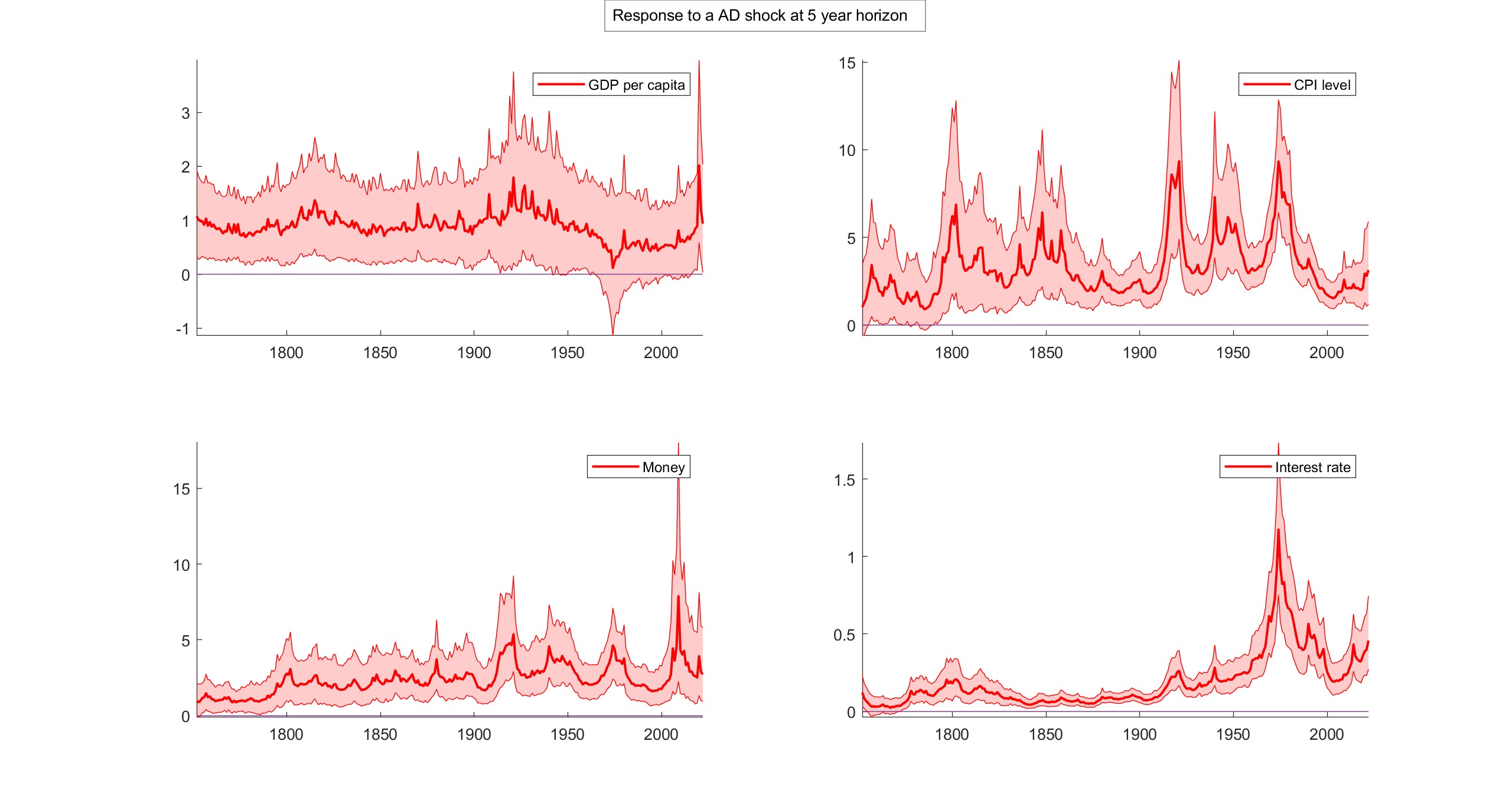} 
\par\end{centering}
{\footnotesize Notes: The pink shaded areas represent the 68 percent
credible sets for the IRFs and the red lines depict the median IRFs
using sign restrictions. }{\footnotesize\par}
\end{figure}

\begin{figure}[H]
\caption{IRF: aggregate supply}

\begin{centering}
\includegraphics[width=0.95\textwidth]{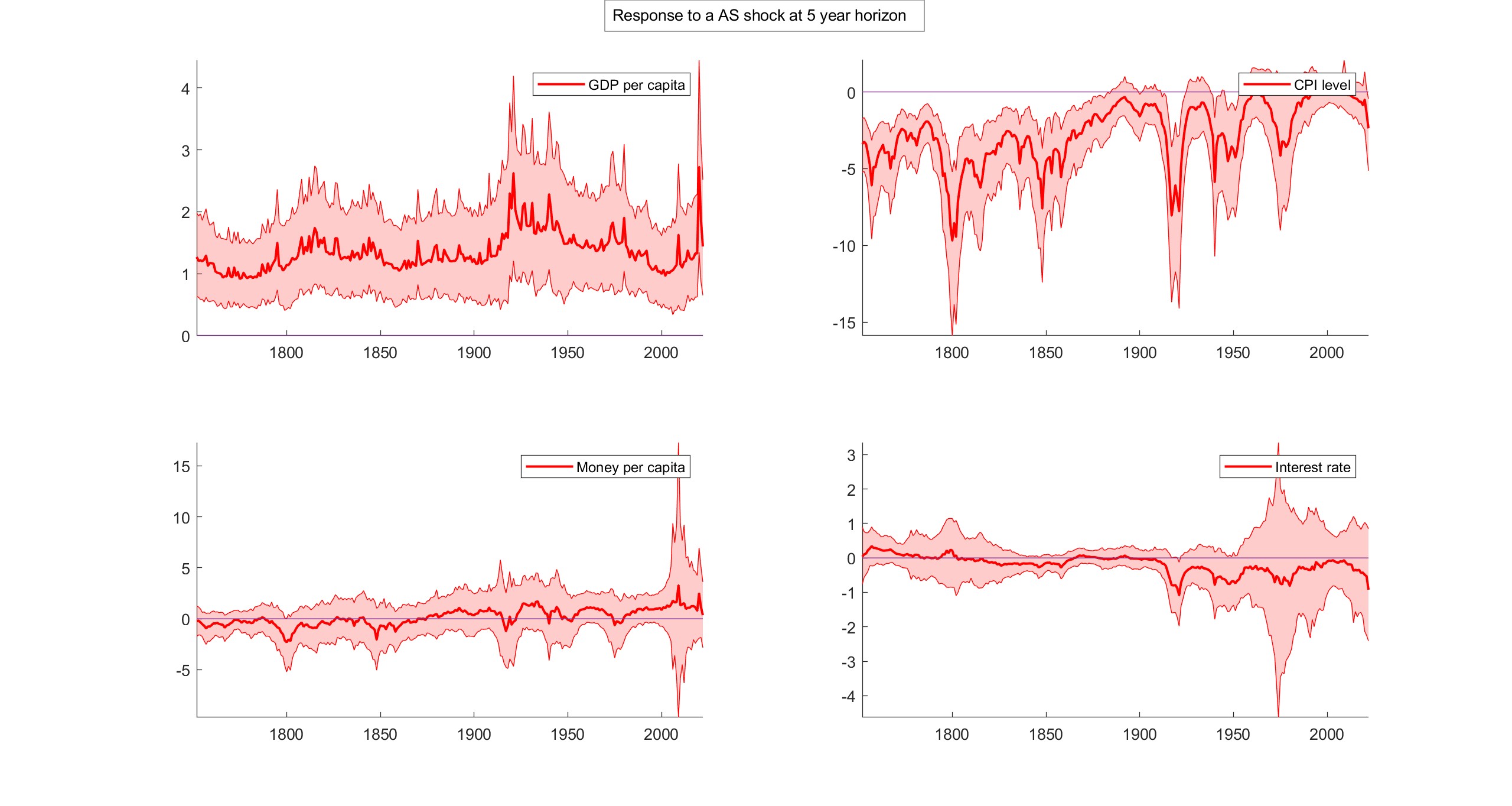} 
\par\end{centering}
{\footnotesize Notes: The pink shaded areas represent the 68 percent
credible sets for the IRFs and the red lines depict the median IRFs
using sign restrictions. }{\footnotesize\par}
\end{figure}

\begin{figure}[H]
\caption{IRF: money demand}

\begin{centering}
\includegraphics[width=1\textwidth]{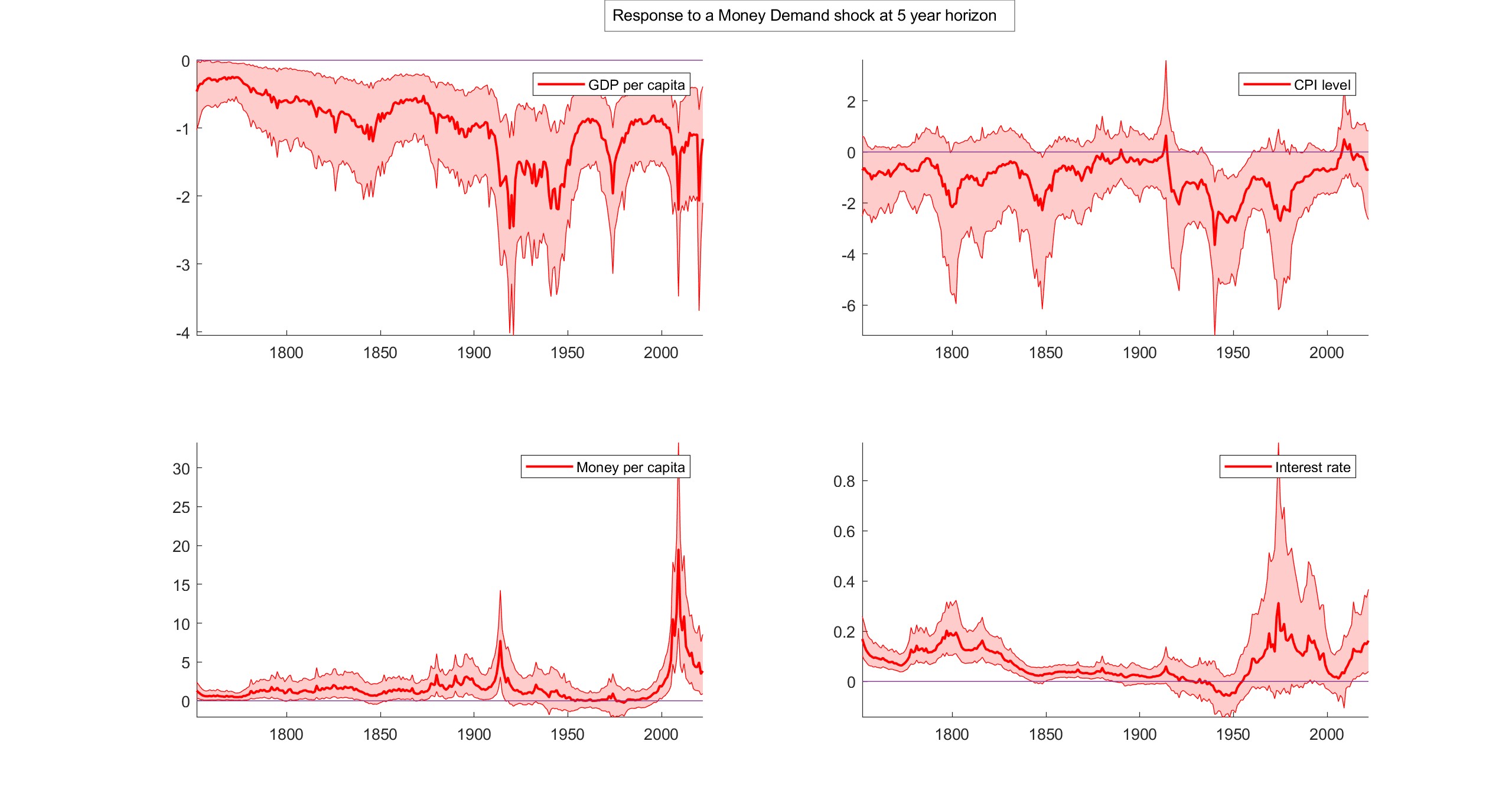} 
\par\end{centering}
{\footnotesize Notes: The pink shaded areas represent the 68 percent
credible sets for the IRFs and the red lines depict the median IRFs
using sign restrictions. }{\footnotesize\par}
\end{figure}

\begin{figure}[H]
\caption{IRF: money supply}

\begin{centering}
\includegraphics[width=1\textwidth]{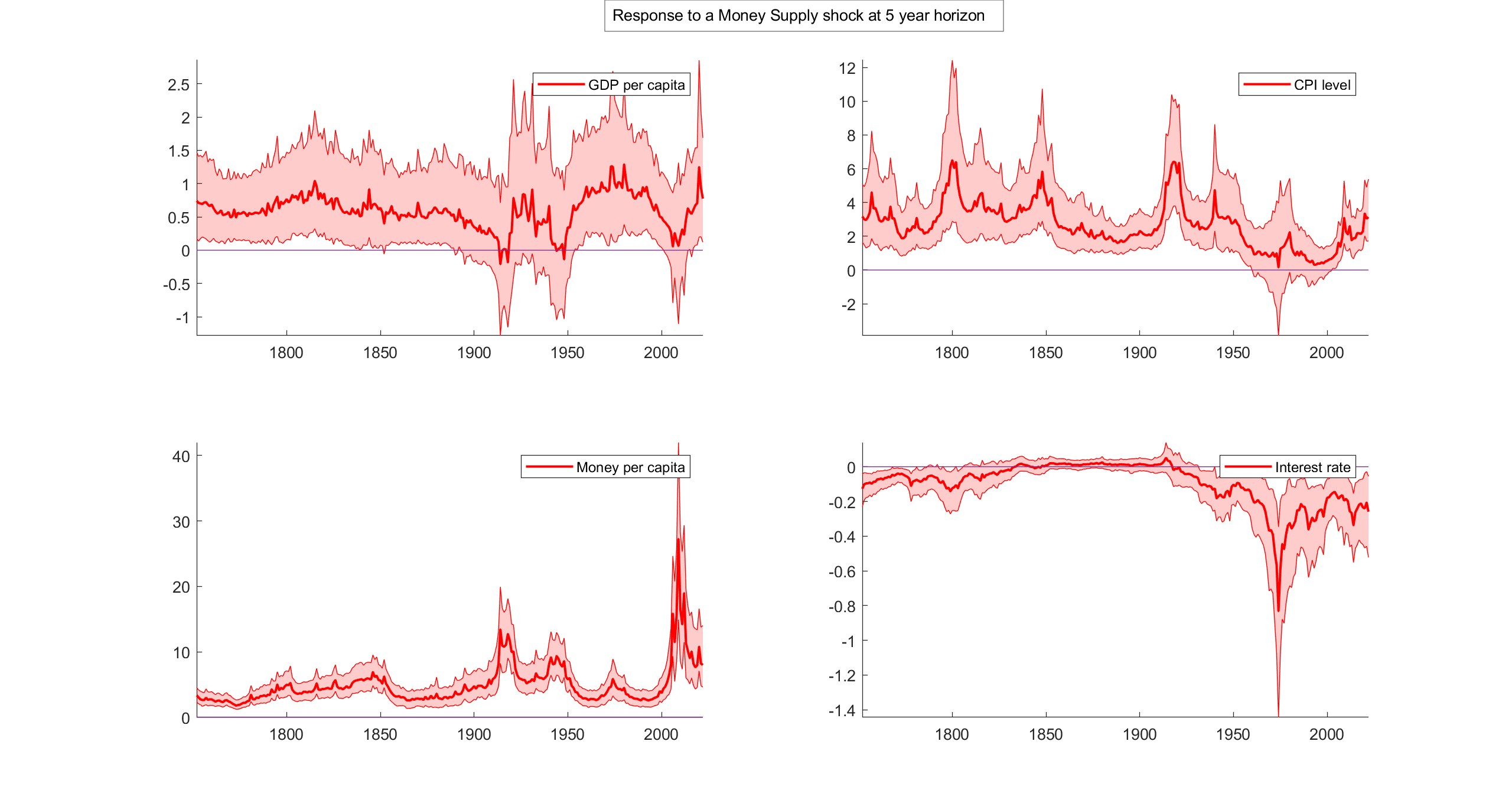} 
\par\end{centering}
{\footnotesize Notes: The pink shaded areas represent the 68 percent
credible sets for the IRFs and the red lines depict the median IRFs
using sign restrictions. }{\footnotesize\par}
\end{figure}

\begin{figure}[H]
\caption{FEVD: aggregate supply}

\begin{centering}
\includegraphics[width=1\textwidth]{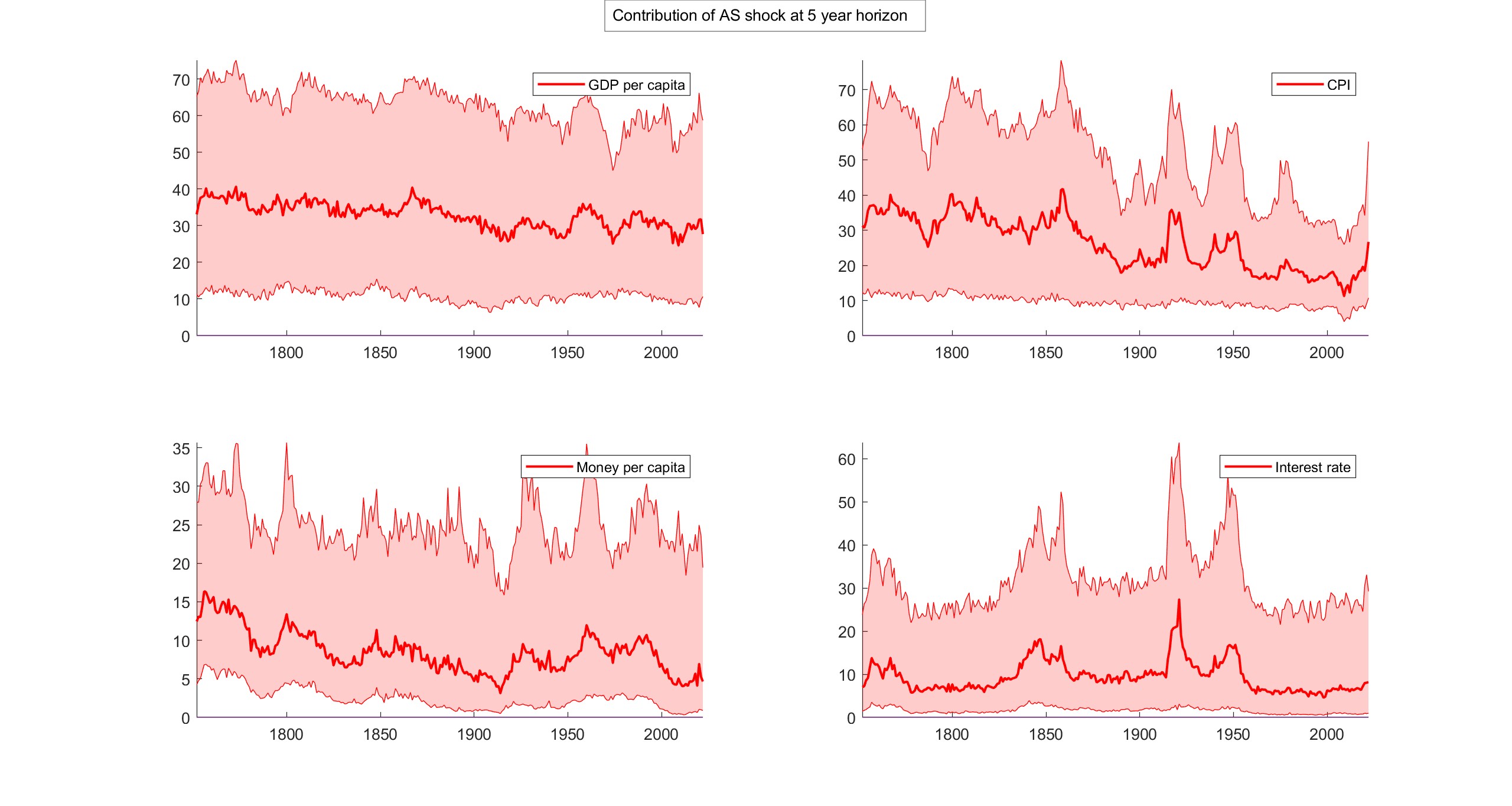} 
\par\end{centering}
{\footnotesize Notes: The pink shaded areas represent the 68 percent
credible sets and the red lines depict the medians. }{\footnotesize\par}
\end{figure}

\begin{figure}[H]
\caption{FEVD: money demand}

\begin{centering}
\includegraphics[width=1\textwidth]{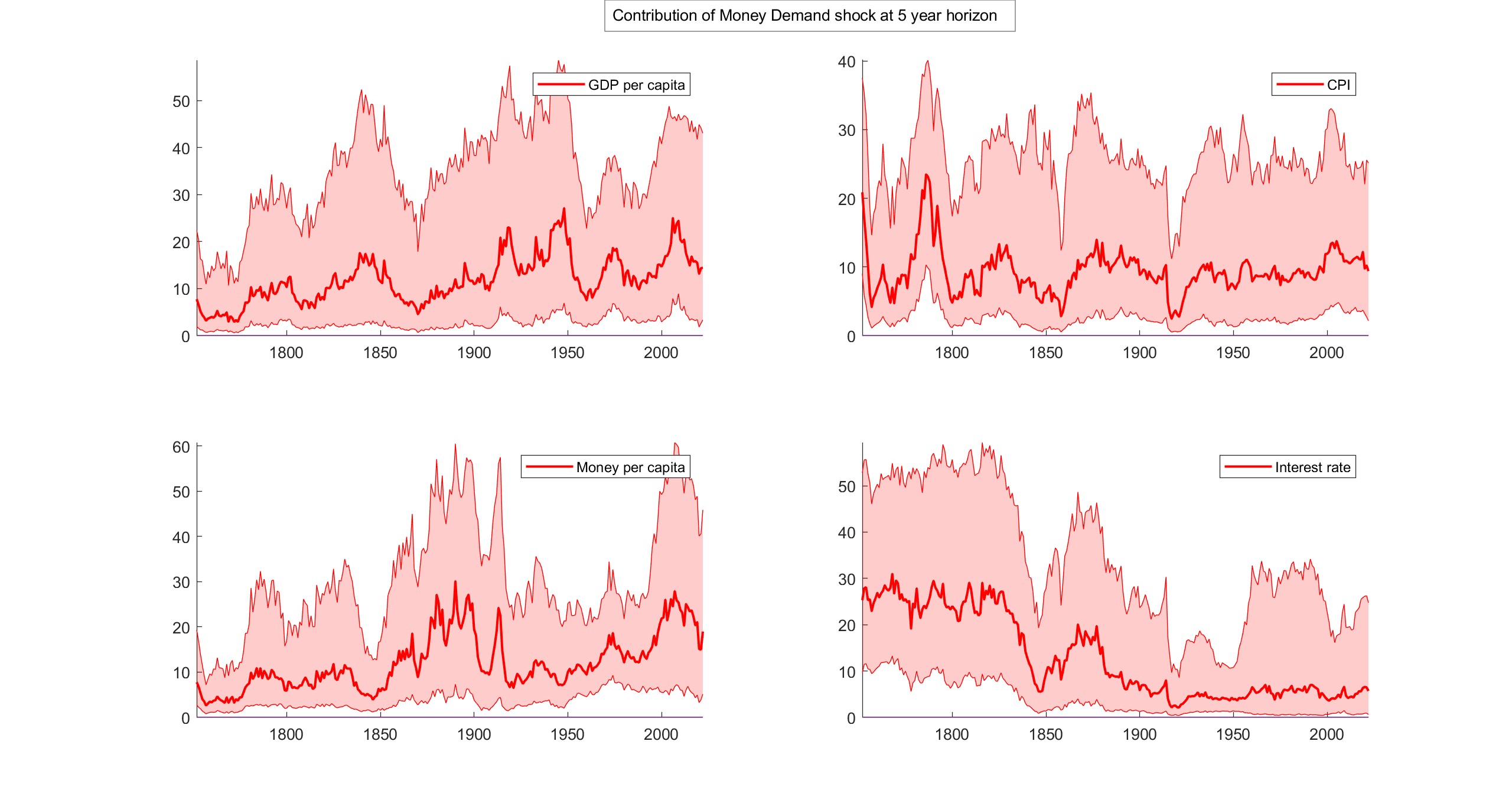} 
\par\end{centering}
{\footnotesize Notes: The pink shaded areas represent the 68 percent
credible sets and the red lines depict the medians. }{\footnotesize\par}
\end{figure}

%\begin{figure}[H] % it's already in the main text
%\caption{FEVD: money supply}
%
%\begin{centering}
%\includegraphics[width=1\textwidth]{fev_ms} 
%\par\end{centering}
%Notes: 
%\end{figure}

\subsubsection{Dynamic Factor Model}

\label{dfm} 
\begin{figure}[H]
\caption{Main IRFs from DFM model\protect}
\label{fig:dfm} 
\begin{centering}
\includegraphics[width=1\textwidth]{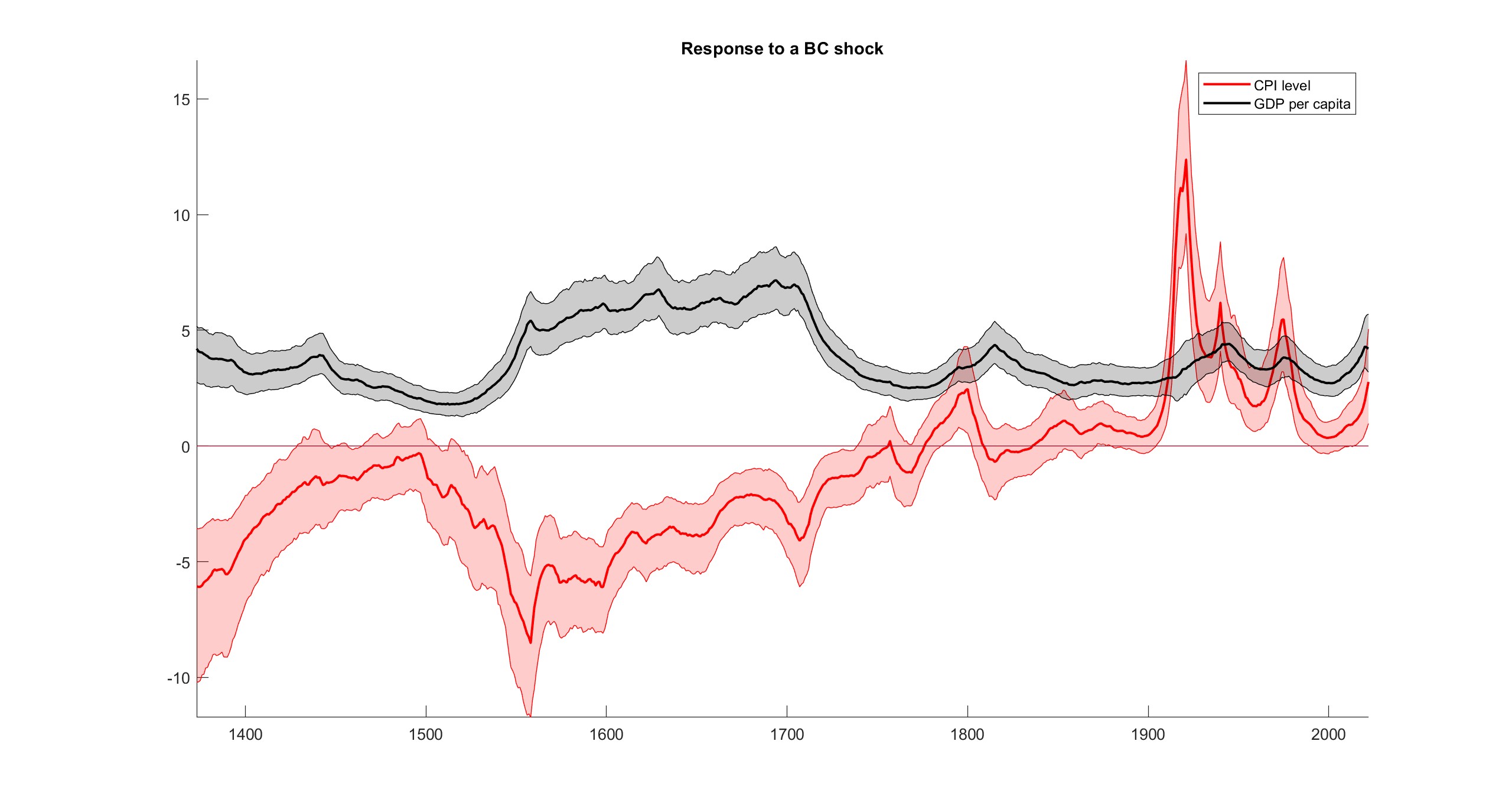} 
\par\end{centering}
{\footnotesize Notes: The pink shaded area represents the 68 percent
credible set and the red line depicts the median IRFs of CPI produced
by a DFM. The grey shaded area and the black line display the equivalent
quantities for the GDP per capita. }{\footnotesize\par}
\end{figure}

\begin{figure}[H]
\caption{Estimated Natural Rate\protect}
\label{fig:rstar} 
\begin{centering}
\includegraphics[width=1\textwidth]{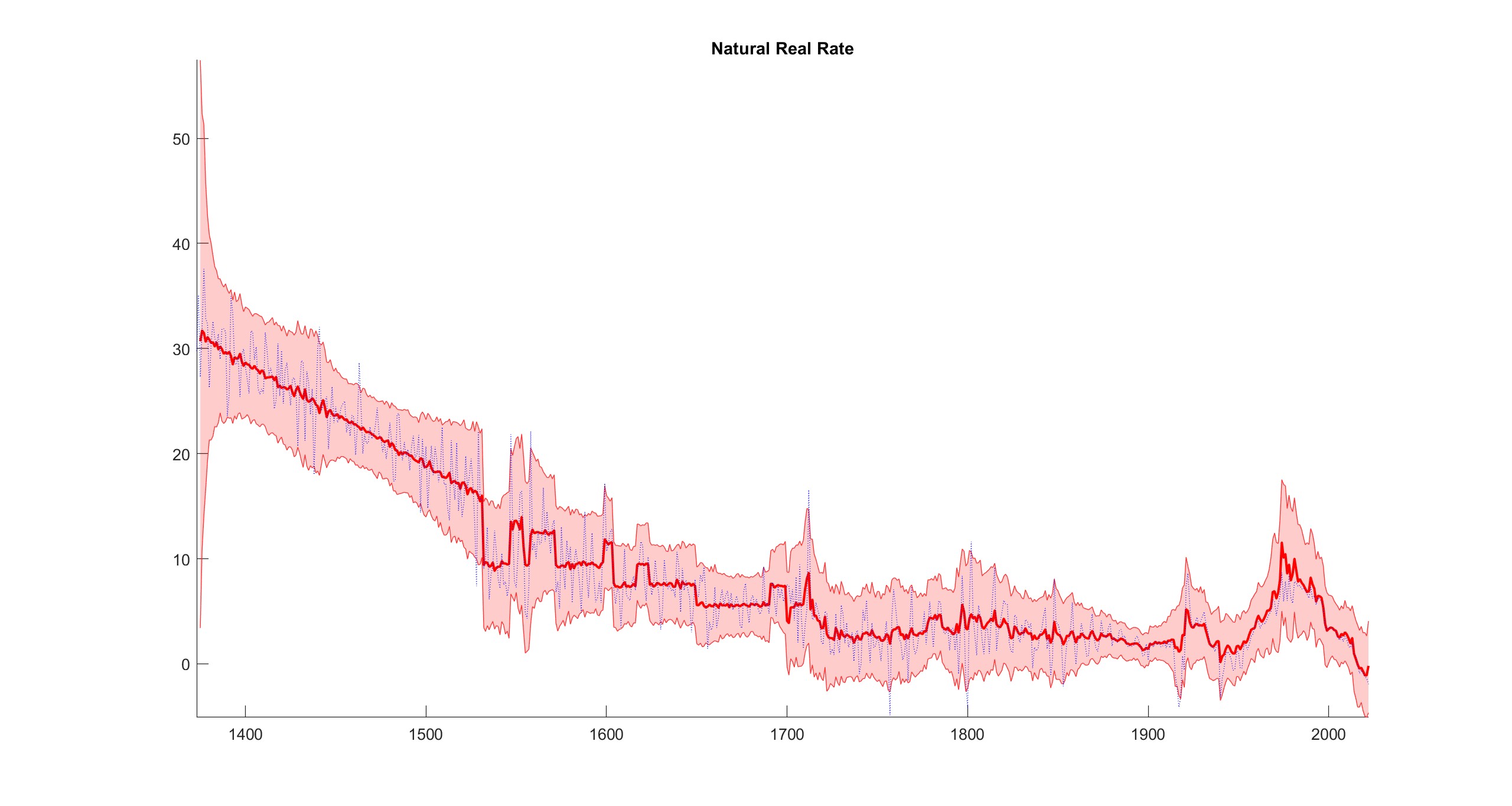} 
\par\end{centering}
{\footnotesize Notes: The pink shaded area represents the 68 percent
credible set and the red line is the median natural rate of interest. }{\footnotesize\par}
\end{figure}

Alternatively, we address measurement error by estimating
a Dynamic Factor Model (DFM). Figure \ref{fig:dfm} shows that IRFs
are similar to baseline results. A by-product of this approach is
an estimate of the natural rate of interest. Following \citet{lubik2015calculating},
we estimate r-star as the 5-year-ahead forecast of the observed real
interest rate. Figure \ref{fig:rstar} shows this estimate for the
UK over seven centuries. In line with \citet{Schmelzing2020boe},
our estimate exhibits a declining trend throughout the early centuries
of the sample, followed by a long period of relative stability. This
stability was disrupted in the late nineteenth century, with the natural
rate peaking at 11.5\% in 1974.

%(Figure \ref{fig:dfm}) 
%\item natural rate of interest estimation (Figure \ref{fig:rstar}) 
%\end{itemize}

\paragraph{Estimation and specification}

We model the dynamics of $11$ variables denoted by $\underset{M\times1}{X_{t}}$
using a Dynamic Factor model with time-varying parameters that allows
for an unbalanced panel. These variables are: real wages, GDP deflator,
CPI, exports, commodity prices, GDP, GDP per capita, M4, shares prices,
the dollar rate and the real interest rate.\footnote{The real interest rate is obtained from \citet{Schmelzing2020boe}
and the other variables are obtained from \citet{Ryland2018} as before.} Variables are included in the model in first differences, with the
exception of the interest rate.

The observation equation of the model is defined as: 
\begin{align*}
X_{t} & =BF_{t}+v_{t}\\
v_{t} & =\rho v_{t-1}+\sigma_{t}^{\frac{1}{2}}e_{t}\\
ln(\sigma_{t}) & =ln(\sigma_{t-1})+g_{t}
\end{align*}
$F_{t}$ denotes $k$ common factors in the dataset $X_{t}$, and
the idiosyncratic components $e_{t}$ accounts for measurement error.

The transition equation is: 
\begin{align*}
F_{t} & =\beta_{t}F_{t-1}+A_{t}^{-1}H_{t}^{\frac{1}{2}}\epsilon_{t}\\
\tilde{\beta}_{t} & =\tilde{\beta}_{t-1}+u_{\beta,t}\\
h_{t} & =h_{t-1}+u_{h,t}\\
a_{t} & =a_{t-1}+u_{a,t}
\end{align*}

The model is estimated using an MCMC algorithm. Priors for $var(u_{\beta,t})$,
$var(u_{h,t})$ and $var(g_{t})$ are based on previous studies such
as \citet{Cogley_Sargent_2005} like in the VAR. 
\end{document}